\documentclass[sigconf]{acmart}
\usepackage[linesnumbered,ruled,vlined]{algorithm2e}
\usepackage{algorithmic}
\usepackage{array}
\newcolumntype{P}[1]{>{\centering\arraybackslash}p{#1}}
\newcolumntype{M}[1]{>{\centering\arraybackslash}m{#1}}
\usepackage{multirow}
\usepackage{balance}
\usepackage{todonotes}
\usepackage{booktabs,subcaption,amsfonts,dcolumn}
\usepackage{paralist}
\newtheorem{definition}{Definition}
\newcommand{\argminD}{\arg\!\min} 
\usepackage{graphicx}
\newcommand{\model}{\textbf{\textit{w}}}

\usepackage{xcolor}

\AtBeginDocument{%
  \providecommand\BibTeX{{%
    \normalfont B\kern-0.5em{\scshape i\kern-0.25em b}\kern-0.8em\TeX}}}

\setcopyright{acmcopyright}
\copyrightyear{2018}
\acmYear{2018}
\acmDOI{XXXXXXX.XXXXXXX}

\acmConference[Conference 'XX]{Make sure to enter the correct
  conference title from your rights confirmation emai}
\acmISBN{978-1-4503-XXXX-X/18/06}

\begin{document}

\title{More is Better (Mostly): On the Backdoor Attacks in Federated Graph Neural Networks}

\author{Jing Xu}
\email{j.xu-8@tudelft.nl}
\affiliation{%
  \institution{Delft University of Technology}
  \country{The Netherlands}
}

\author{Rui Wang}
\email{r.wang-8@tudelft.nl}
\affiliation{%
  \institution{Delft University of Technology}
  \country{The Netherlands}
}

\author{Stefanos Koffas}
\email{S.Koffas@tudelft.nl}
\affiliation{%
  \institution{Delft University of Technology}
  \country{The Netherlands}
}

\author{Kaitai Liang}
\email{Kaitai.Liang@tudelft.nl}
\affiliation{%
  \institution{Delft University of Technology}
  \country{The Netherlands}
}

\author{Stjepan Picek}
\email{picek.stjepan@gmail.com}
\affiliation{%
  \institution{Radboud University \& Delft University of Technology}
  \country{The Netherlands}
}

\renewcommand{\shortauthors}{Jing and Rui, et al.}

\begin{abstract}
Graph Neural Networks (GNNs) are a class of deep learning-based methods for processing graph domain information. GNNs have recently become a widely used graph analysis method due to their superior ability to learn representations for complex graph data.
However, due to privacy concerns and regulation restrictions, centralized GNNs can be difficult to apply to data-sensitive scenarios.
Federated learning (FL) is an emerging technology developed for privacy-preserving settings when several parties need to train a shared global model collaboratively.
Although several research works have applied FL to train GNNs (Federated GNNs), there is no research on their robustness to backdoor attacks.

This paper bridges this gap by conducting two types of backdoor attacks in Federated GNNs: centralized backdoor attacks (CBA) and distributed backdoor attacks (DBA).
Our experiments show that the DBA attack success rate is higher than CBA in almost all evaluated cases.
For CBA, the attack success rate of all local triggers is similar to the global trigger even if the training set of the adversarial party is embedded with the global trigger.
To further explore the properties of two backdoor attacks in Federated GNNs, we evaluate the attack performance for a different number of clients, trigger sizes, poisoning intensities, and trigger densities.
Moreover, we explore the robustness of DBA and CBA against one defense. We find that both attacks are robust against the investigated defense, necessitating the need to consider backdoor attacks in Federated GNNs as a novel threat that requires custom defenses.
\end{abstract}




\maketitle

\section{Introduction}
\label{sec:introduction}

Graph Neural Networks, which generalize traditional deep neural networks (DNNs) to graph data, pave a new way to effectively learn representations for complex graph-structured data~\cite{DBLP:journals/tnn/WuPCLZY21}. Due to their strong representation learning capability, GNNs have demonstrated remarkable performance in various domains, e.g., drug discovery~\cite{xiong2019pushing, lim2019predicting}, finance~\cite{wang2019semi, cheng2022financial}, social networks~\cite{fan2019graph, guo2020deep}, and recommendation systems~\cite{yin2019deeper, fan2019metapath}. 
Usually, GNNs are trained through centralized training. However, because of privacy concerns, regulatory restrictions, and commercial competition, GNNs can also face challenges when centrally trained.
For example, the financial institution may utilize GNN as a fraud detection model, but they can only have transaction data of its registered users (no data of other users because of privacy concerns). Thus, the model is not effective for other users. 
Similarly, in a drug discovery industry that applies GNNs, pharmaceutical research institutions can dramatically benefit from other institutions' data, but they cannot disclose their private data for commercial reasons~\cite{hefedgraphnn}.

Federated Learning is a distributed learning paradigm that works on isolated data. In FL, clients can collaboratively train a shared global model under the orchestration of a central server while keeping the data decentralized~\cite{kairouz2019advances, mcmahan2017communication}. As such, FL is a promising solution for training GNNs over isolated graph data, and there are already some works utilizing FL to train GNNs~\cite{hefedgraphnn, zhang2021federated, lalitha2019peer}, which we denote as \textit{Federated GNNs}.

Although FL has been successfully applied in diverse domains, e.g., computer vision~\cite{liu2020fedvision, liu2020federated} or language processing~\cite{zhu2020empirical, hard2018federated}, there could be malicious clients among millions of clients, leading to various adversarial attacks~\cite{bagdasaryan2020backdoor, fang2020local}. In particular, limited access to local clients' data due to privacy concerns or regulatory constraints may facilitate backdoor attacks on the global model trained in FL. 
A backdoor attack is a type of poisoning attack that manipulates part of the training dataset with a specific pattern (trigger) such that the model trained on the manipulated dataset will misclassify the testing dataset with the same trigger pattern~\cite{liu2017trojaning}. 

Backdoor attacks on FL have been recently studied~\cite{bagdasaryan2020backdoor, bhagoji2019analyzing, xie2019dba}. However, these attacks are applied in federated learning on Euclidean data, e.g., images and words.  
The backdoor trigger generation methods and injecting position are different between graph data and images/words~\cite{DBLP:conf/wisec/XuXP21}. In particular, in~\cite{xie2019dba}, the authors split a square-shaped trigger placed in the top left corner of an image into four parts so that four malicious clients use each part in their poisoned datasets. When the training ends, the adversary concatenates these parts to form a global trigger in the image's upper left corner that activates the backdoor. This is impossible in GNNs as the data is not Euclidean, and there is no position that we can exploit. Also, defenses like FoolsGold~\cite{fung2018mitigating} filter out clients that use similar updates as malicious. This can be effective for Euclidean data that use parts of the trigger in similar positions but may not be effective in GNNs. Indeed, the graph data is not Euclidean, and different partial triggers vary the graph structure resulting in non-aligned updates. 
Additionally, intensive research has been conducted on backdoor attacks in GNNs~\cite{DBLP:conf/sacmat/ZhangJWG21, DBLP:conf/uss/XiPJ021, DBLP:conf/wisec/XuXP21}. However, these works focus on GNN models in centralized training. In federated learning, the malicious updates will be weakened in the aggregation function. Finally, there can be more than one malicious client, while in centralized GNNs, there is only one client. 
Thus, we should expect different behavior of backdoor attacks in Federated GNNs. Then, it is crucial to investigate if existing countermeasures that have been tested mostly with Euclidean data are still effective for backdoor attacks in Federated GNNs to understand how to deploy trustworthy AI systems.

This paper conducts two backdoor attacks in FL: centralized backdoor attacks (CBA) and distributed backdoor attacks (DBA)~\cite{xie2019dba}.
In CBA, the attacker embeds the same global trigger in all adversarial clients, while in DBA, the adversary decomposes the global trigger into several local triggers and embeds them in different malicious clients. 
In DBA, we assume two attack scenarios - honest majority and malicious majority, to explore the impact of the percentage of malicious clients on the attack. Our work focuses on the cross-silo federated learning setting, and our main contributions are:
\begin{itemize}
    \item We explore two types of backdoor attacks in Federated GNNs. Based on the experiments, we find that the DBA on Federated GNNs is more effective or (at least) similar to the CBA. To the best of our knowledge, this paper is the first work studying backdoor attacks in Federated GNNs.
    \item We find that in the CBA, although the adversarial local model is implanted with the global trigger, the final global model can also attain promising attack performance with any local trigger. Since this phenomenon is inconsistent with the related works, we provide further experiments to explain it.
    \item We observe that in most cases, local triggers in DBA can achieve similar attack performance to the global trigger, which is different from the findings for the DBA in Convolutional Neural Networks (CNNs).
    \item We run experiments for both types of attacks, varying the trigger size, poisoning intensity, and trigger density, and show that the trigger size has more impact than the poisoning intensity.
    \item We explore the robustness of DBA and CBA against one defense, i.e., FoolsGold. We find both attacks are evasive to FoolsGold, while CBA can even obtain a higher attack success rate, but the testing accuracy degrades.

\end{itemize}

\section{Background}
\label{sec:background}

\subsection{Federated Learning}

Federated Learning enables $n$ clients to train a global model $\model$ collaboratively without revealing local datasets. Unlike centralized learning, where local datasets must be collected by a central server before training, FL performs training by uploading the weights of local models ($\{\model^i \mid i \in n\}$) to a parametric server. 
Specifically, FL aims to optimize a loss function:
\begin{equation}
    \mathop {\min }\limits_\model \ \ell (\model ) = \sum\limits_{i = 1}^n {\frac{{{k_i}}}{n}{L_i}(} \model ),{L_i}(\model ) = \frac{1}{{{k_i}}}\sum\limits_{j \in {P_i}} {{\ell _j}(\model, x_j)},
\end{equation}
where ${L_i}(\model )$ and $k_i$ are the loss function and local data size of $i$-th client, and $P_i$ refers to the set of data indices with size $k_i$.

At the $t$-th iteration, the training can be divided into three steps: 
\begin{itemize}
    \item \textit{Global model download}. All clients download the global model $\model_t$ from the server.
    \item \textit{Local training}.  Each client updates the global model by training with their datasets: $\model_t^i \leftarrow \model_t^i-\eta\frac{\partial L(\model_t, b)}{\partial \model_t^i} $, where $\eta$ and $b$ refer to learning rate and local batch, respectively. 
    \item \textit{Aggregation}. After the clients upload their local models $\{\model_t^i \mid i \in n\}$, the server updates the global model by aggregating the local models. In this paper, we use the averaging aggregation function: $\model_{t+1} \leftarrow \sum\limits_{i = 1}^n \frac{{{1}}}{n} \model_t^i.$
\end{itemize}

\subsection{Graph Neural Networks}

Recently, Graph Neural Networks (GNNs) have achieved significant success in processing non-Euclidean spatial data, which are very common in many real-world scenarios. Unlike traditional neural networks, e.g., CNNs and Recurrent Neural Networks (RNNs), GNNs work on graph data. 
GNNs take a graph $G=(V,E,X)$ as an input, where $V, E, X$ denote nodes, edges, and node attributes, and learn a representation vector (embedding) for each node $\boldsymbol{v} \in G$, $z_{\boldsymbol{v}}$, or the entire graph, $z_G$. 

Modern GNNs follow a neighborhood aggregation strategy, where one iteratively updates the representation of a node by aggregating representations of its neighbors. After $k$ iterations of aggregation, a node's representation captures both structure and feature information within its $k$-hop network neighborhood~\cite{xu2018powerful}. Formally, the $k$-th layer of a GNN is (e.g., GCN~\cite{kipf2017semi}, GraphSAGE~\cite{DBLP:conf/nips/HamiltonYL17}, and GAT~\cite{velickovic2018graph}):
\begin{equation}
    z_{\boldsymbol{v}}^{(k)} = \sigma(z_{\boldsymbol{v}}^{(k-1)}, AGG(\{ z_{\boldsymbol{u}}^{(k-1)}; \boldsymbol{u} \in \mathcal{N}_{\boldsymbol{v}} \})), \forall k \in [K],
    \label{eqn:2.2-1}
\end{equation}
where $z_{\boldsymbol{v}}^{(k)}$ is the representation of node $\boldsymbol{v}$ computed in the $k$-th iteration. $\mathcal{N}_{\boldsymbol{v}}$ are neighbors of node $\boldsymbol{v}$, and the $AGG(\cdot)$ is an aggregation function that can vary for different GNN models. $z_{\boldsymbol{v}}^{(0)}$ is initialized as node feature, while $\sigma$ is an activation function.
For the graph classification task (considered in this work), the READOUT function pools the node representations for a graph-level representation $z_G$:
\begin{equation}
    z_G = READOUT({z_{\boldsymbol{v}};v \in V}).
    \label{eqn:2.2-2}
\end{equation}
READOUT can be a simple permutation invariant function such as summation or a more sophisticated graph-level pooling function~\cite{DBLP:conf/nips/YingY0RHL18, DBLP:conf/aaai/ZhangCNC18}. 

\subsection{Backdoor Attacks on Federated Learning}

Backdoor attacks aim to make a model misclassify its inputs to a preset-specific label without affecting its original task. Attackers poison the model by injecting triggers into the training data that activate the backdoor in the test phase.
Once activated, the model's output becomes the targeted label pre-specified by the attacker to achieve the malicious intent purpose (such as misclassification). 

Backdoor attacks are common in FL systems with multiple training dataset owners.
Specifically, the adversary $\mathcal{A}$ manipulates one or more local models to obtain poisoned models $\tilde{W}^i$ that are then aggregated into the global model $G_t$ affecting its properties. 
There are two common techniques used in backdoor attacks in FL: 1) data poisoning where $\mathcal{A}$ manipulates local training dataset(s) $D_{local}^i$ used to train the local model~\cite{nguyen2020poisoning, xie2019dba}, and 2) model poisoning where $\mathcal{A}$ manipulates the local training process or the trained local models themselves~\cite{bagdasaryan2020backdoor}. 
In this work, we use data poisoning for our attacks in Federated GNNs as model poisoning requires multiplying large factors to model weights when conducting attacks, which can be detected by traditional byzantine-robust aggregation rules such as Median~\cite{yin2018byzantine} and Krum~\cite{blanchard2017machine}. 





\section{Problem Formulation}
\label{sec:problem}

\subsection{Overview}

FL is a practical choice to push machine learning to users' devices, e.g., smart speakers, cars, and phones. Usually, federated learning is designed to work with thousands or even millions of users without restrictions on eligibility~\cite{bagdasaryan2020backdoor}, opening up new attack vectors. 
As stated in~\cite{bonawitz2019towards}, training with multiple malicious clients is now considered a practical threat by the designers of federated learning.
Because of the data privacy guarantee among the clients in the federated learning, local clients can modify their local training dataset without being noticed. Furthermore, existing federated learning frameworks do not provide a functionality to verify whether the training on local clients has been finished correctly. Consequently, one or more clients can submit their malicious models trained for the assigned task and backdoor functionality.

\subsection{Threat Model}
\label{subsec:threat}

Unlike traditional machine learning benchmarking datasets, graph datasets, and real-world graphs may exhibit non-independent and identical distribution (non-i.i.d) due to factors like structure and feature heterogeneity~\cite{hefedgraphnn}.
Therefore, following the FL assumptions, we assume that graphs among $K$ clients are non-i.i.d. distributed. 
The clients engaging in training can be divided into honest and malicious clients. 
In Table~\ref{tab:setting_k_m},\footnote{Exp. I, Exp. II, Exp. III, and Exp. IV represent the experiments of the honest majority attack scenario, malicious majority attack scenario, the impact of the number of clients, and the impact of the percentage of malicious clients, respectively.} we summarize the settings of different experiments shown in Section~\ref{sec:experiments}. Molecular machine learning is a paramount application in the Federated GNNs, where many small graphs are distributed between multiple institutions~\cite{hefedgraphnn}. Therefore, we run experiments (Exp. I and II) on two molecular datasets, i.e., NCI1 and PROTEINS\_full. For these experiments, we set 5 clients in total because, with more clients, the local dataset of each client becomes very small, resulting in severe overfitting for the local models. Similar settings and phenomena can also be found in prior works on Federated GNNs~\cite{hefedgraphnn}. The choice of small datasets may be a limitation of our work, but real-world cross-silo settings could involve only a few different organizations (from two to one hundred)~\cite{kairouz2019advances}.
Besides the molecular domain, substantial attention has also been given to Federated GNNs in real-world financial scenarios~\cite{zhang2021federated, wang2022review}. In such scenarios, clients can be different organizations, e.g., banks, and a GNN model is trained on siloed data, leading to a  cross-silo federated learning setting~\cite{kairouz2019advances}. As shown in Exp. III and IV, we assume $10$, $20$, and $100$ clients for a synthetic dataset, i.e., TRIANGLES, which is a realistic real-world cross-silo scenario~\cite{shejwalkar2022back}.

\begin{table}[!htb]
\footnotesize
 \centering
 \caption{Summary of the experimental setting ($K$: number of clients, $M$: number of malicious clients).}
\begin{tabular}{cccc} 
 \hline
 Experiment & Dataset & $K$ & $M$ \\
 \hline
 Exp. I & NCI1, PROTEINS\_full, TRIANGLES & $5$ & $2$ \\
 \hline
 Exp. II & NCI1, PROTEINS\_full, TRIANGLES & $5$ & $3$ \\
 \hline
 \multirow{2}{*}{Exp. III} & \multirow{2}{*}{TRIANGLES} & $10$ & $4, 6$ \\
  & & $20$ &$8, 12$ \\
 \hline
 Exp. IV & TRIANGLES & $100$ & $5, 10, 15, 20$ \\
 \hline
 Prior work~\cite{hefedgraphnn} & Molecules & $4$ & $0$ \\
 \hline
\end{tabular}
\label{tab:setting_k_m}
\end{table}

All clients strictly follow the FL training process, but the malicious client(s) will inject graph trigger(s) into their training graphs. 
We also assume the server is conducting model aggregation correctly. 
Our primary focus is to investigate backdoor attack effectiveness on Federated GNNs, so we adopt two backdoor attack methods as defined below (the definitions of the local trigger and global trigger used in these two attacks are also given).

\begin{definition}[Local Trigger \& Global Trigger.] The local trigger is the specific graph trigger for each malicious client in DBA. The global trigger is the combination of all local triggers.\footnote{Since it is an NP-hard problem to decompose a graph into subgraphs~\cite{dasgupta2008algorithms}, we first generate local triggers and then compose them to get the global trigger used in CBA.}
\end{definition}

\begin{definition}[Distributed Backdoor Attack (DBA).] There are multiple malicious clients, and each of them has its local trigger. 
Each malicious client injects its local trigger into its training dataset. All malicious clients have the same backdoor task.
An adversary $\mathcal{A}$ conducts DBA by compromising at least two clients in FL.
\end{definition}

\begin{definition}[Centralized Backdoor Attack (CBA).] A global trigger consisting of local triggers is injected into one client's local training dataset. An adversary $\mathcal{A}$ conducts CBA by usually compromising only one client in FL.
\end{definition}

\textbf{Adversary's capability.} 
We assume the adversary $\mathcal{A}$ can corrupt $M$ ($M \leq K$) clients to perform DBA.
We perform a complete attack in every round, i.e., a poisoned local dataset is used by malicious clients in every round, following the attack setting in~\cite{xie2019dba}.
The adversary cannot impact the aggregation process on the central server nor the training or model updates of other clients.

\textbf{Adversary's knowledge.} We assume that the adversary $\mathcal{A}$ knows the compromised clients' training dataset. In this context, the adversary can generate local triggers as described in Section~\ref{section:4.2}.
Additionally, we follow the original assumptions of FL. The number of clients participating in training, model structure, aggregation strategy, and a global model for each iteration is revealed to all clients, including malicious clients. 

\textbf{Adversary's goal.} Unlike some non-targeted attacks~\cite{poursaeed2018generative} aiming to deteriorate the accuracy of the model, the backdoor attacks studied in this paper aim to make the global model misclassify the backdoored data samples into specific pre-determined labels (i.e., target label $y_t$) without affecting the accuracy on clean data. 

In distributed backdoor attacks, each malicious client injects its local trigger into its local training dataset to poison the local model. Therefore, DBA can fully leverage the power of FL in aggregating dispersed information from local models to train a poisoned global model. Assuming there are $M$ malicious clients in DBA, each has its local trigger. 
Each malicious client $i$ in DBA independently implements a backdoor attack on its local model. The adversarial objective for each malicious client $i$ is:

\begin{equation} 
\label{eqn:3-2}
\begin{split}
{w_{t}^{i}}^* &= \argminD_{w_{t}^i} (\sum_{j\in D_{trigger}^i}\mathit{\ell} (w_{t-1}^i(\Phi(x_j^i, \kappa^i), y_t)) \\
 &+ \sum_{j \in D_{clean}^i}\ell(w_{t-1}^i(x_j^i), y_j^i)), \forall{i \in [M]},
\end{split}
\end{equation}
where the poisoned training dataset $D_{trigger}^i$ and clean training dataset $D_{clean}^i$ satisfy $D_{trigger}^i \cup D_{clean}^i = D_{local}^i$ and $D_{trigger}^i \cap D_{clean}^i = \varnothing$. $D_{local}^i$ is the local training dataset of client $i$. $\Phi$ is the function that transforms the clean data with a non-target label into poisoned data using a set of trigger generation parameters $\kappa^i$. 
In this paper, $\kappa^i$ consists of trigger size $s$, trigger density $\rho$, and poisoning intensity $r$: $\kappa = \left \{s, \rho, r\right \}$. 

\textbf{Trigger Size $s$}: the number of nodes of a local graph trigger. Here, we set the trigger size $s$ to be the $\gamma$ fraction of the graph dataset's average number of nodes. Note that this does not violate our threat model (the adversary does not have access to the whole dataset) as the average number of nodes in the local dataset is similar to the number of the whole dataset.\\
\textbf{Trigger Density $\rho$}: the complexity of a local graph trigger, which ranges from $0$ to $1$, and is used in the Erdős-Rényi (ER) model to generate the graph trigger.\\
\textbf{Poisoning Intensity $r$}: the ratio that controls the percentage of backdoored training dataset among the local training dataset.

Unlike DBA with multiple malicious clients, there is only one malicious client in CBA.\footnote{In practice, the centralized attack can poison more than one client with the same global trigger, as mentioned in~\cite{bagdasaryan2020backdoor}. Here, we assume there is one malicious client}
CBA is conducted by embedding a global trigger into a malicious client's training dataset. The global trigger is a graph consisting of local trigger graphs used in DBA, as explained further in Section~\ref{section:4-1}. 
Thus, the adversarial objective of the attacker $k$ in round $t$ in CBA is:

\begin{equation} 
\label{eqn:3-1}
\begin{split}
{w_{t}^{k}}^* &= \argminD_{w_{t}^k} (\sum_{j\in D_{trigger}^k}\mathit{\ell} (w_{t-1}^k(\Phi(x_j^k, \kappa), y_t)) \\
 &+ \sum_{j \in D_{clean}^k}\ell(w_{t-1}^k(x_j^k), y_j^k)),
\end{split}
\end{equation}
where $\kappa$ is the combination of $\kappa^i$. 
Utilizing the power of FL in message passing from local models to the global model, the global model is supposed to inherit the backdoor functionality.


\section{Backdoor Attacks against Federated GNNs}
\label{sec:backdoor}


\subsection{General Framework}
\label{section:4-1}

\begin{figure}[!htb]
\centering
\includegraphics[scale=.25, page=1]{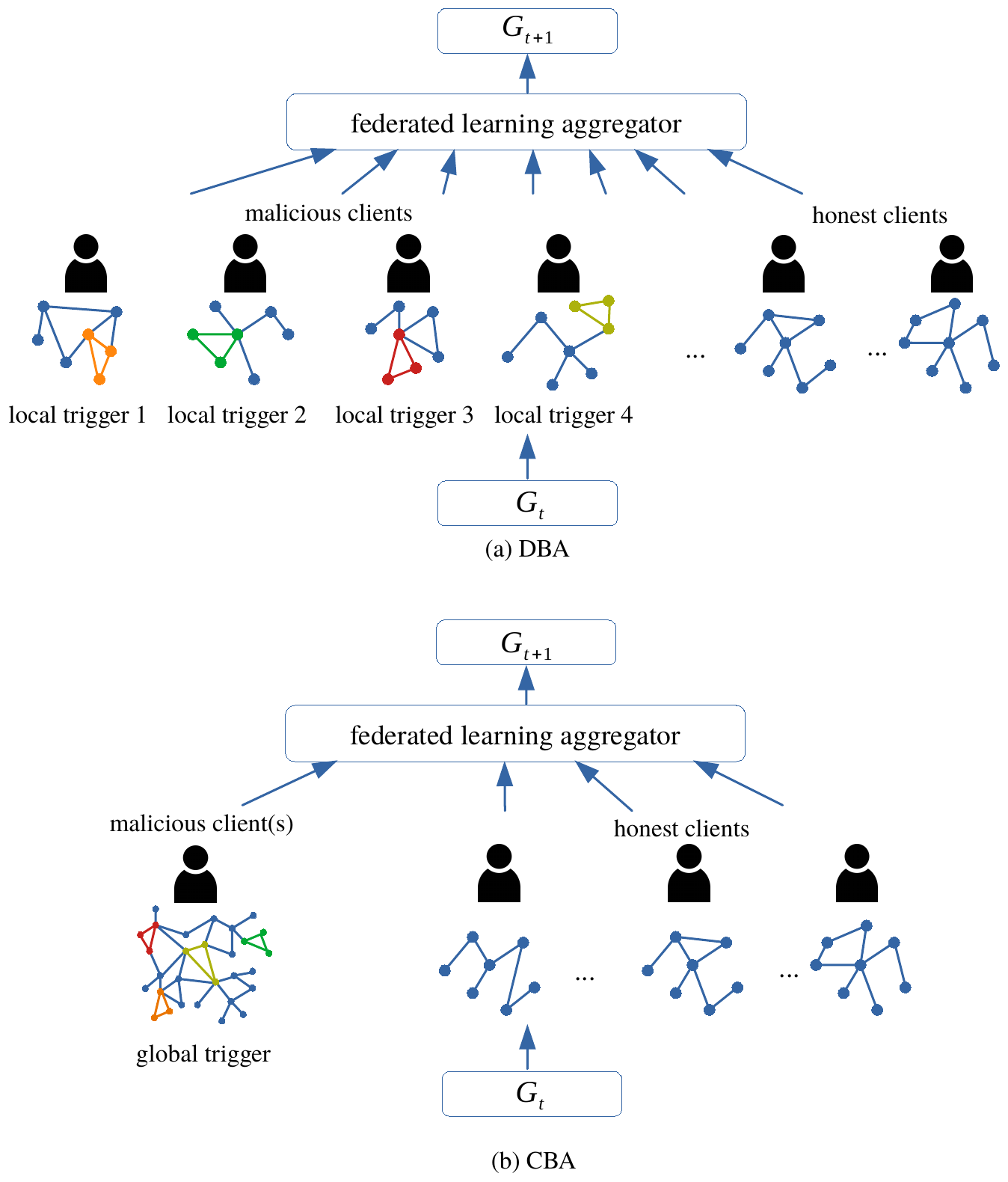}
\caption{\label{fig:attack_framework}Attack Framework.}
\end{figure}

We focus on subgraph-based (data poisoning) backdoor attacks and the graph classification task.
Attackers can perform DBA or CBA as shown in Figure~\ref{fig:attack_framework}.
In DBA, multiple malicious clients engage in attacking, and they inject local triggers into corresponding malicious clients' local training datasets.
CBA is conducted with one malicious client, whose training data is poisoned with the global trigger that consists of the local triggers used in DBA.
We describe the notations used throughout the paper in Table~\ref{tab:notations} in Appendix~\ref{sec:appendix-notation}.

\textbf{Distributed Backdoor Attack.}
For DBA in Federated GNNs, we assume there are $M$ ($M \leq K$) malicious clients among $K$ clients, as shown in Figure~\ref{fig:attack_framework}(a).
Each malicious client embeds its local training dataset with a specific graph trigger to poison its local model. 
For instance, in Figure~\ref{fig:attack_framework}(a), each malicious client has a local trigger highlighted by a specific color (i.e., orange, green, red, yellow).\footnote{Although we use the triangle as the graph trigger for each malicious client, in practice, the local triggers are more complex and different from each other.}
In this paper, we did not use the same local trigger for different malicious clients in DBA as 
it would mean poisoning intensity for this specific local trigger is increasing, but simultaneously, the total trigger pattern activating the backdoor is reduced. We evaluated this setting by running some additional experiments, and we found the attack under this setting is not stronger than the current setting (i.e., different local triggers).
Through training with these poisoned training datasets, the poisoned local models are uploaded to the server to update the global model.
The final adversarial goal is to use the global trigger to attack the global model.
Algorithms~\ref{alg:DBA} and~\ref{alg:client_update} illustrate the distributed backdoor attack in Federated GNNs.
We first split the clients into two groups, the honest ($C_h$) and the malicious one ($C_m$) (line 2, Algorithm~\ref{alg:DBA}).
In each round, each client updates its weights through local training (line 13,  Algorithm~\ref{alg:DBA}), and finally, the global server aggregates local models' weights to update the global model through averaging (line 15,  Algorithm~\ref{alg:DBA}). 

The local training for every client is described in Algorithm~\ref{alg:client_update}. If the client is malicious (line 2,  Algorithm~\ref{alg:client_update}), the local training dataset will be backdoored (line 4, Algorithm~\ref{alg:client_update}) with the local trigger (line 3, Algorithm~\ref{alg:client_update}. As mentioned in Section~\ref{subsec:threat}, all the local triggers form the global trigger (line 5, Algorithm~\ref{alg:client_update}). 

We conduct experiments for the malicious majority and honest majority settings to explore the impact of different percentages of malicious clients on the attack success rate. We provide additional motivation for the malicious majority setting in Section~\ref{sec:related}.

\textbf{Centralized Backdoor Attack.} 
Unlike DBA conducted with multiple malicious clients, CBA performs the attack with only one malicious client. CBA is a general approach in a centralized learning scenario. For example, in image classification, the attacker poisons the training dataset with a trigger so that the model misclassifies the data sample with the same trigger into the attacker-chosen label. As shown in Figure~\ref{fig:attack_framework}(b), the malicious client embeds its training dataset with the global trigger highlighted by four colors. 
This global trigger consists of local triggers used in DBA, as shown in Line 5 of Algorithm~\ref{alg:client_update}.
Specifically, the attacker in CBA embeds its training data with four local patterns, together constituting a complete global pattern as the backdoor trigger.\footnote{Here, the four colors are only used to denote four trigger patterns.} 

To compare the attack performance between the distributed backdoor attack and centralized backdoor attack in Federated GNNs, we need to make sure the trigger pattern in CBA is the union set of local trigger patterns in DBA.
We can use two strategies: 1) first generate local triggers in DBA and then combine them to get the global trigger, or 2) first generate a global trigger in CBA and then divide it into $M$ local triggers. We utilize the first strategy as it is an NP-hard problem to divide a graph into several subgraphs~\cite{dasgupta2008algorithms}. Thus, in different attack scenarios (i.e., honest majority or malicious majority attack scenarios), the CBA performance is different since the global trigger has been changed due to the different number of malicious clients.

\begin{algorithm}[!htb]
\footnotesize
\SetAlgoLined
\caption{Distributed Backdoor Attacks in Federated GNNs} \label{alg:DBA}
\SetKwInput{KwInput}{Input}
\SetKwInput{KwOutput}{Output}
\DontPrintSemicolon
    \KwInput{
    Dataset $D$, Target label $y_t$\\}
    \KwOutput{
    Backdoored Global model $G_{t+1}$, global trigger $t_{global}$\\
    }
    
    \SetKwFunction{FMain}{DBA()}
    
    \SetKwProg{Fn}{Function}{:}{}
    \Fn{\FMain}{
        $C_h, C_m \leftarrow ClientSplit(Clients)$\;
        $D_{local}, D_{test} \leftarrow DataSplit(D)$\;
        $t_{global} \leftarrow \varnothing$\;
        \textbf{Server executes:}\;
        initialize $G_0$, $f=False$\;
        \ForEach{round $t=0, 1, 2, ...$}{
            \ForEach{client $k \in (C_h \cup C_m)$}{
                $w_t^k = G_t$ \;
                \If{$k \in C_m$}{
                $f = True$\;
                }
                $w_{t+1}^k \leftarrow ClientUpdate(k, w_t^k, f, t_{global})$\;
            }
           $G_{t+1} \leftarrow \sum_{k=1}^{K}\frac{w_{t+1}^k}{K}$
        }
    }
    \textbf{End Function}\;
    \textbf{return} $G_{t+1}, t_{global}$
\end{algorithm}

\begin{algorithm}[!htb]
\footnotesize
\SetAlgoLined
\caption{ClientUpdate} \label{alg:client_update}
\SetKwInput{KwInput}{Input}
\SetKwInput{KwOutput}{Output}
\DontPrintSemicolon
    \KwInput{
    Client $k$, Local training dataset $D_{local}$, Current global model $w$, flag $f$, global trigger $t_{global}$\\}
    \KwOutput{
    Updated model $w$\\}
    \SetKwFunction{FMain}{ClientUpdate()}
    \SetKwProg{Fn}{Function}{:}{}
    \Fn{\FMain}{
        \If{f is $True$}{
            $t_{local} \leftarrow GenerateTrigger(s, \rho)$\;
            $D_{local}\leftarrow BackdoorDataset(D_{local}, t_{local}, y_t)$\;
            $t_{global} = t_{global} \cup t_{local}$\;
            }
        $\mathcal{B} \leftarrow$ (split $D_{local}$ into batches of size B)\;
        \ForEach{local epoch $i$ from $1$ to $E$}{
            \ForEach{$b \in \mathcal{B}$}{
                $w \leftarrow w-\eta \bigtriangledown l(w, b)$\;
            }
        }
    }
    \textbf{End Function}\;
    \textbf{return} $w$
\end{algorithm}

\subsection{Backdoored Data Generation}
\label{section:4.2}

We adopt the Erdős-Rényi (ER) model~\cite{Gilbert1959} to generate triggers (function \textit{GenerateTrigger} in Algorithm~\ref{alg:client_update}) as it is more effective than the other methods (e.g., Small World model~\cite{watts1998collective} or Preferential Attachment model~\cite{barabasi1999emergence})~\cite{DBLP:conf/sacmat/ZhangJWG21}.
In particular, \emph{GenerateTrigger} (line 3 in Algorithm~\ref{alg:client_update}), creates a random graph of $s$ nodes. An edge between a pair of nodes in this graph is generated with probability $\rho$. 

Backdoored data is generated (line 4 in Algorithm~\ref{alg:client_update}) through the following process.
We sample subsets of the local training datasets (with non-target labels) with proportion $r$, and the rest are saved as clean datasets. For each sampled data, we inject a trigger into it by sampling $s$ (trigger size) nodes from the graph uniformly at random and replacing their connection with that in the trigger graph. Additionally, the attacker re-labels the
sampled data with an attacker-chosen target label. The backdoored data is composed of the dataset with trigger and the original clean dataset.

\section{Experiments}
\label{sec:experiments}

\subsection{Experimental Setting}
\label{section:5.1}


We implemented FL algorithms using the PyTorch framework.
All experiments were run on a server with 2 Intel Xeon CPUs, one NVIDIA 1080 Ti GPU with 32GB RAM. 
Each experiment was repeated ten times to obtain the average result.
Our code is blinded for review but will be made public.

\textbf{Datasets.} We run experiments on three publicly available datasets: 
two molecular structure datasets - NCI1~\cite{morris2020tudataset}, PROTEINS\_full~\cite{borgwardt2005protein},
and one synthetic dataset - TRIANGLES~\cite{knyazev2019understanding}, which is a multi-class dataset.
Table~\ref{table:dataset_statistics} in Appendix~\ref{sec:appendix-dataset-statistic} provides more information about these datasets.


\textbf{Dataset splits.} For each dataset, we randomly sample $80\%$ of the data instances as the training dataset and the rest as the test dataset. To simulate non-i.i.d. training data and supply each participant with an unbalanced sample from each class, we further split the training dataset into $K$ parts following the strategy in~\cite{fang2020local} with hyperparameter $0.5$ for TRIANGLES (10 classes) and hyperparameter $0.7$ for other datasets (2 classes).
In this paper, apart from Appendix~\ref{section:analysis-of-factors} where we analyze the effect of trigger factors, we set trigger factors as follows: $\gamma = 0.2$, $\rho = 0.8$, and $r = 0.2$. As we show in Appendix~\ref{section:analysis-of-factors}, these hyperparameters yield an effective attack. By choosing them, we model a strong adversary that helps in evaluating the attack's behavior in the worst-case scenario.

\textbf{Models and metrics.} In our experiments, we use three state-of-the-art GNN models: GCN~\cite{kipf2017semi}, GAT~\cite{velickovic2018graph}, and GraphSAGE~\cite{DBLP:conf/nips/HamiltonYL17}.

We use the \textit{attack success rate (ASR)} to evaluate the attack effectiveness. 
We embed the testing dataset with local triggers or the global trigger and then calculate the ASR of the global model on the poisoned testing dataset. We only embed the testing dataset of the non-target label with triggers to avoid the influence of the original label.
The ASR measures the proportion of trigger-embedded inputs 
that are misclassified by the backdoored GNN into the target class $y_t$ chosen by the adversary. 
The trigger-embedded inputs are \[D_{g_t} = \left \{ (G_{1, g_{t}}, y_1), (G_{2, g_{t}}, y_2),  \ldots, (G_{n, g_{t}}, y_n) \right \}.\]
Here, $g_t$ is the graph trigger, $\left \{G_{1, g_t}, G_{2, g_t} \ldots, G_{n, g_t}\right \}$ is the test dataset embedded with graph trigger $g_t$, and ${y_1, y_2, \ldots, y_n}$ is the label set.
Formally, ASR is defined as:
\begin{align*}
 Attack \: Success \: Rate &= \frac{\sum_{i=1}^{n} \mathbb{I}(G_{backdoor}(G_{i, g_{t}}) = y_t)}{n},
 \label{eqn:4.1-1}
\end{align*}
where $\mathbb{I}$ is an indicator function 
and $G_{backdoor}$ refers to the backdoored global model.
Here, the graph trigger $g_t$ can be local triggers or a global trigger.

%

\subsection{Backdoor Attack Results}

We evaluate multiple-shot attack~\cite{bagdasaryan2020backdoor}, which means that the attackers perform attacks in multiple rounds, and the malicious updates are accumulated to achieve a successful backdoor attack. We do not evaluate the single-shot attack~\cite{bagdasaryan2020backdoor} because the multi-shot is stealthier~\cite{pillutla2019robust}.
The multi-shot attack does not require multiplying large factors to model weights when conducting the attack, while the single-shot needs to multiply large factors to maintain the effectiveness of backdoor attacks, which can be filtered out or detected by traditional anomaly detection-based approaches such as Krum~\cite{blanchard2017machine}. Since our main goal is conducting backdoor attacks on FL, we chose a multiple-shot attack with a high attack success rate and stealthiness.
As mentioned in Section~\ref{subsec:threat}, we perform a complete attack in every round, showing the difference between DBA and CBA in a shorter time~\cite{xie2019dba}.

To explore the impact of different percentages of malicious clients on the attack performance, we evaluate the honest majority and malicious majority attack scenarios according to the percentage of malicious clients among all clients. Specifically, we set two and three malicious clients among five clients for the honest majority and malicious majority attack scenarios, respectively. 

In our experiments, we evaluate the ASR of CBA and DBA with the global trigger and local triggers.
The goal is to explore:
\begin{itemize}
    \item In CBA, whether the ASR of local triggers can achieve similar performance to the global trigger even if the centralized attacker would embed a global trigger into the model.
    \item In DBA, whether the ASR of the global trigger is higher than all local triggers even if the global trigger never actually appears in any local training dataset, as mentioned in~\cite{xie2019dba}.
\end{itemize}

\begin{figure}[htbp]
     \centering
     \begin{subfigure}[b]{0.48\textwidth}
         \centering
         \includegraphics[width=\textwidth]{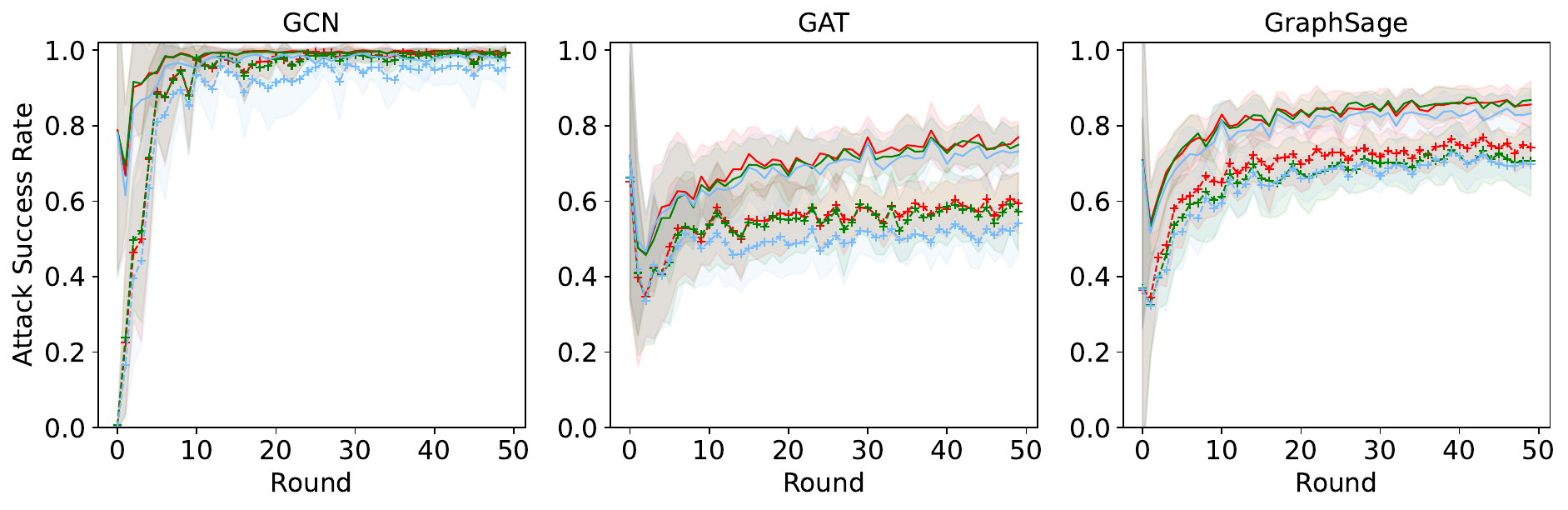}
         \caption{NCI1}
         \label{fig:bk_honest_majority_a}
     \end{subfigure}
     \begin{subfigure}[b]{0.48\textwidth}
         \centering
         \includegraphics[width=\textwidth]{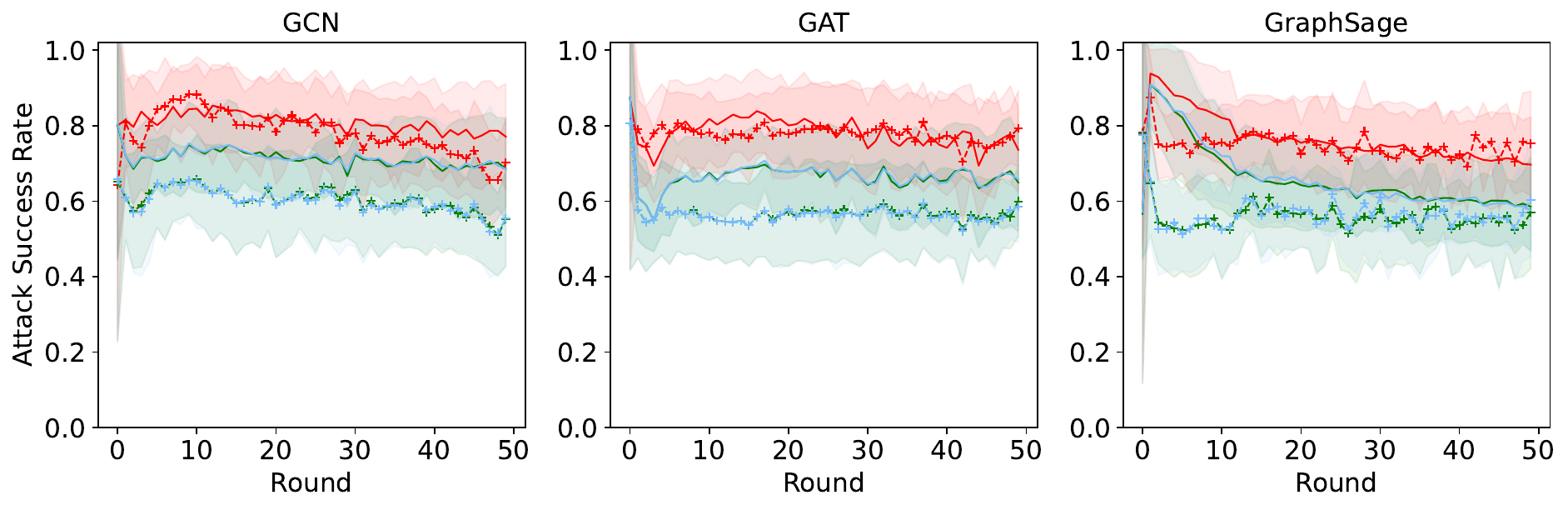}
         \caption{PROTEINS\_full}
         \label{fig:bk_honest_majority_b}
     \end{subfigure}
     \begin{subfigure}[b]{0.48\textwidth}
         \centering
         \includegraphics[width=\textwidth]{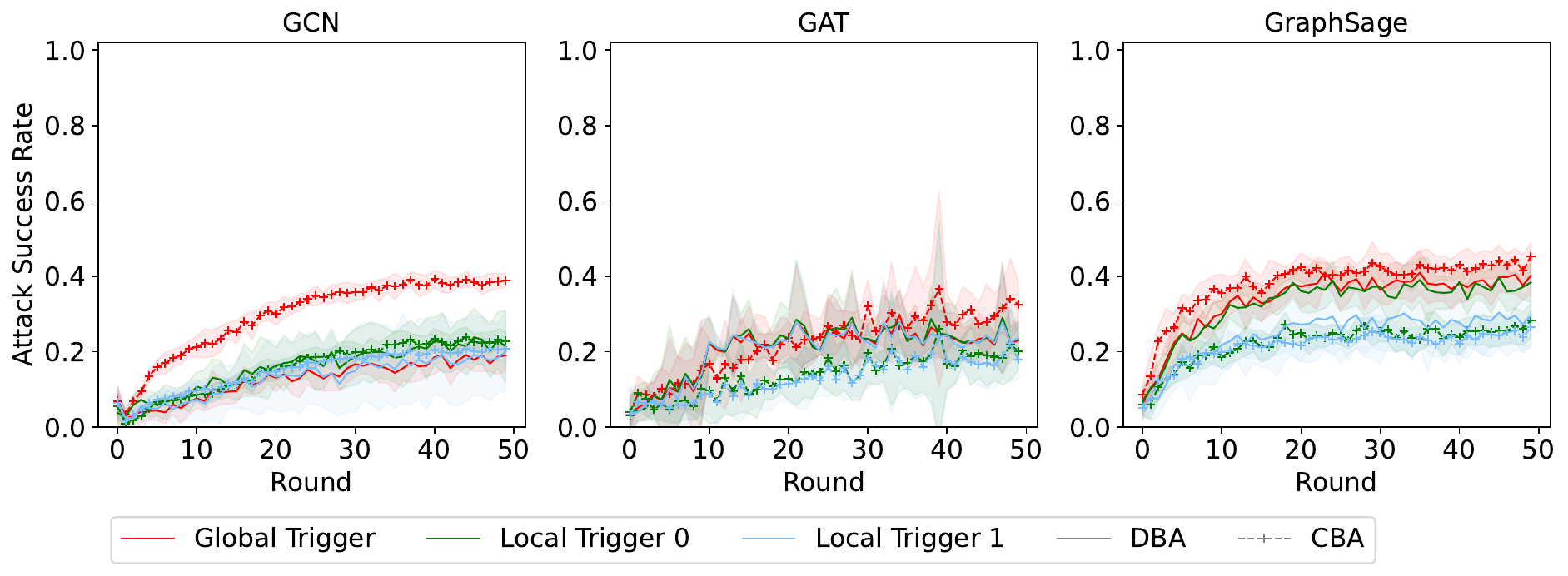}
         \caption{TRIANGLES}
         \label{fig:bk_honest_majority_c}
     \end{subfigure}
     \caption{Backdoor attack results in the honest majority attack scenario.}
     \label{fig:bk_honest_majority}
\end{figure}

\begin{figure}[htbp]
     \centering
     \begin{subfigure}[b]{0.475\textwidth}
         \centering
         \includegraphics[width=1\textwidth]{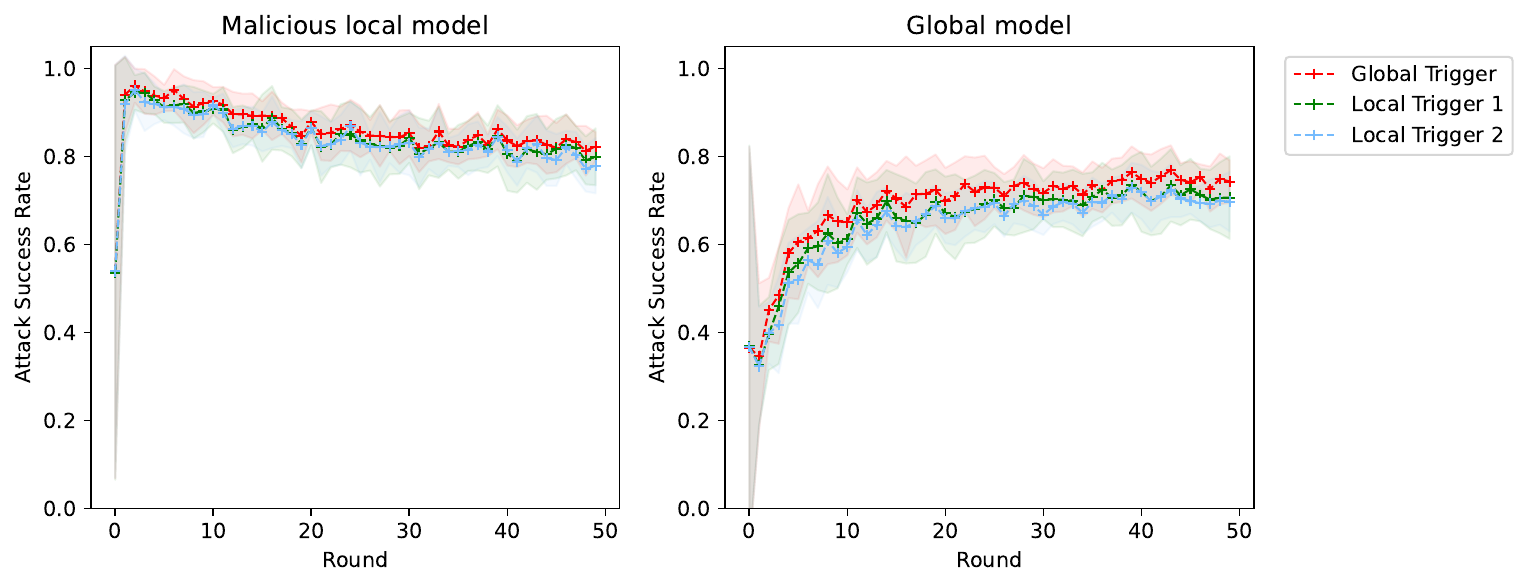}
         \caption{Honest majority attack scenario}
         \label{fig:CBA_analysis_a}
     \end{subfigure}
     \hfill
     \begin{subfigure}[b]{0.475\textwidth}
         \centering
         \includegraphics[width=1\textwidth]{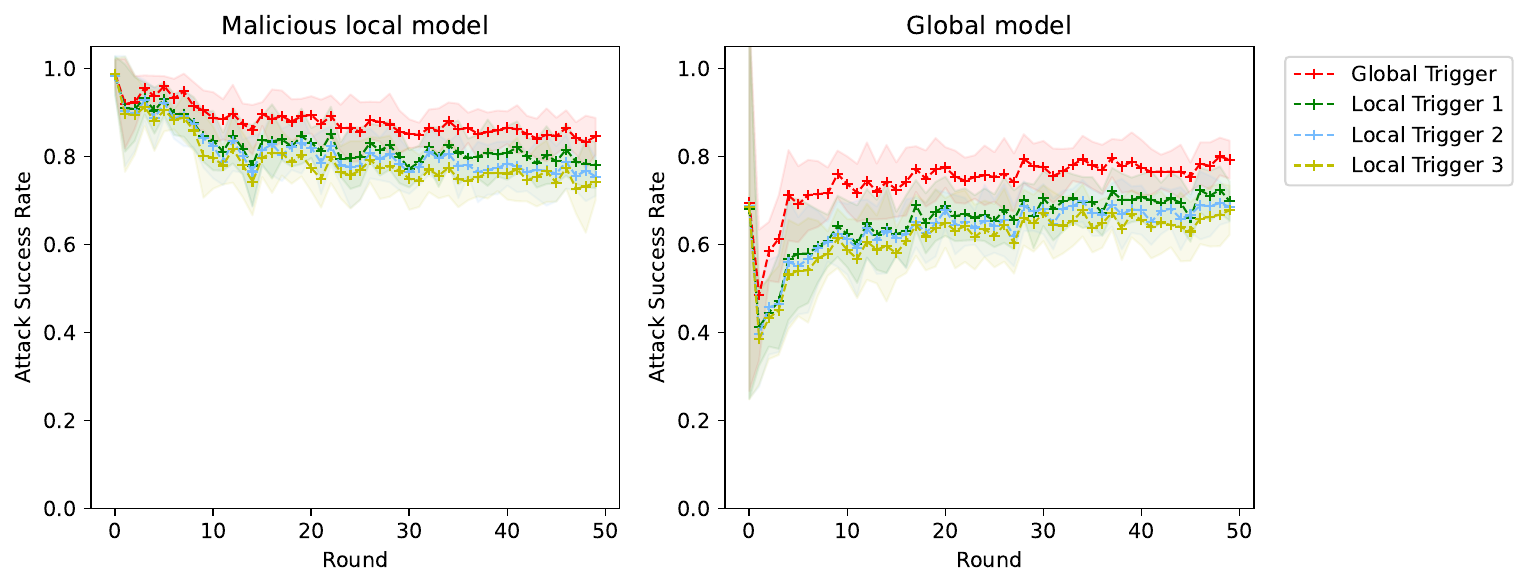}
         \caption{Malicious majority attack scenario}
         \label{fig:CBA_analysis_b}
     \end{subfigure}
     \caption{Centralized backdoor attack results on the malicious local model and global model with different triggers.}
     \label{fig:CBA_analysis}
\end{figure}

\textbf{Honest Majority Attack Scenario.}
\label{exp:honest_majority}
The attack results of CBA and DBA in the honest majority attack scenario are shown in Figure~\ref{fig:bk_honest_majority}. Notice that the DBA ASR with a specific trigger is always higher than or at least similar to that of CBA with the corresponding trigger.  
For example, in Figure~\ref{fig:bk_honest_majority_a} (the result for the GAT model), the DBA ASR with the global trigger is higher than CBA with a global trigger. The only exception happens for GCN on TRIANGLES. 
We also find that the ASR of the two attacks in TRIANGLES is significantly lower than the other two datasets but still higher than random guessing. 
The TRIANGLES is a multi-class dataset containing complex data relations. Thus, more information needs to be encoded in each model's weights for the class features compared to the other datasets. As a result, there is not enough remaining space to learn our triggers easily.
In most results on NCI1 and PROTEINS\_full, there is an initial drop in the attack success rate for both DBA and CBA, resulting from the high local learning rate of honest clients~\cite{bagdasaryan2020backdoor}.
\textit{Based on the result for CBA, surprisingly, the ASR of all local triggers can be as high as the global trigger even if the centralized attacker embeds the global trigger into the model, which is inconsistent with the behavior in~\cite{xie2019dba}.} 
We analyze it through further experiments shown in Figure~\ref{fig:CBA_analysis}.

Moreover, the results for the PROTEINS\_full dataset show that \textit{in DBA, the attack success rate of the global trigger is higher than (or at least similar to) any local trigger, even if the global trigger never actually appears in any local training dataset.}
This indicates that the high attack success rate of the global trigger does not require the same high attack success rate of local triggers. 
However, for the other two datasets (NCI1 and TRIANGLES), the attack success rate of the global trigger is close to all local triggers (except the result of GraphSage on TRIANGLES). This indicates that in some cases, the local trigger embedded in local models can successfully transfer to the global model so that once any local trigger is activated, the global model will misclassify the data sample into the attacker-chosen target label. This phenomenon is not consistent with the observations in~\cite{xie2019dba} as in Euclidean data, most locally triggered images are similar to the clean image, but any (small) change in the structure of a graph will result in a significant dissimilarity.  

\begin{figure}[htbp]
     \centering
     \begin{subfigure}[b]{0.48\textwidth}
         \centering
         \includegraphics[width=\textwidth]{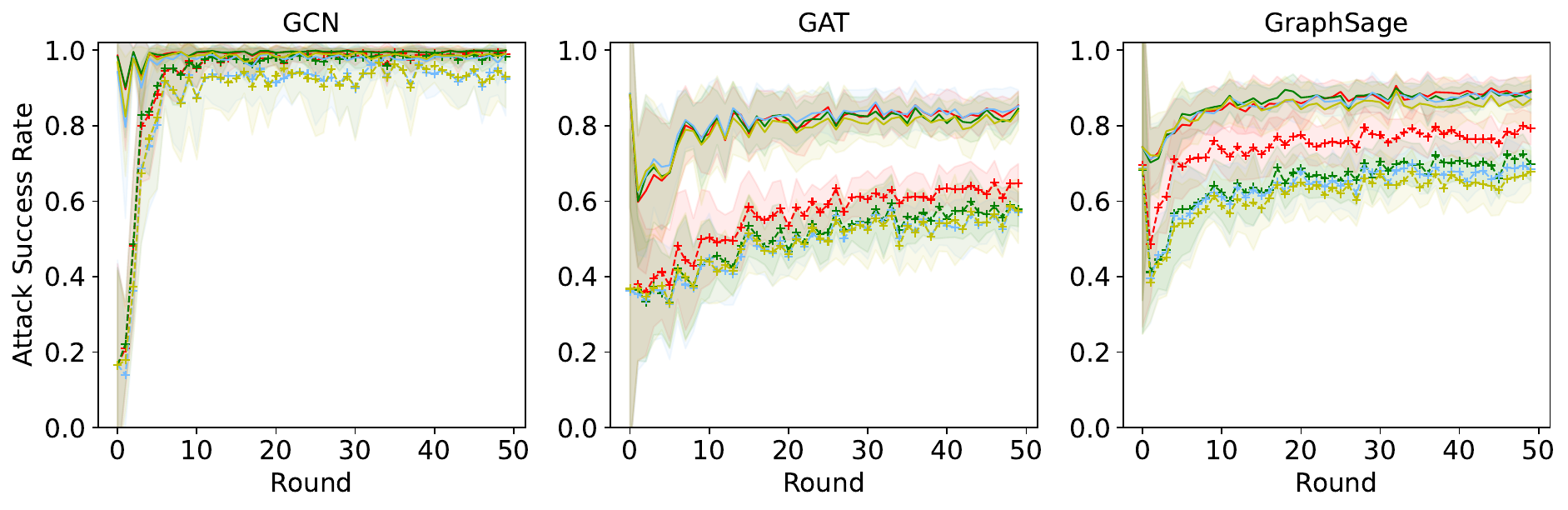}
         \caption{NCI1}
         \label{fig:bk_malicious_majority_a}
     \end{subfigure}
     \hfill
     \begin{subfigure}[b]{0.48\textwidth}
         \centering
         \includegraphics[width=\textwidth]{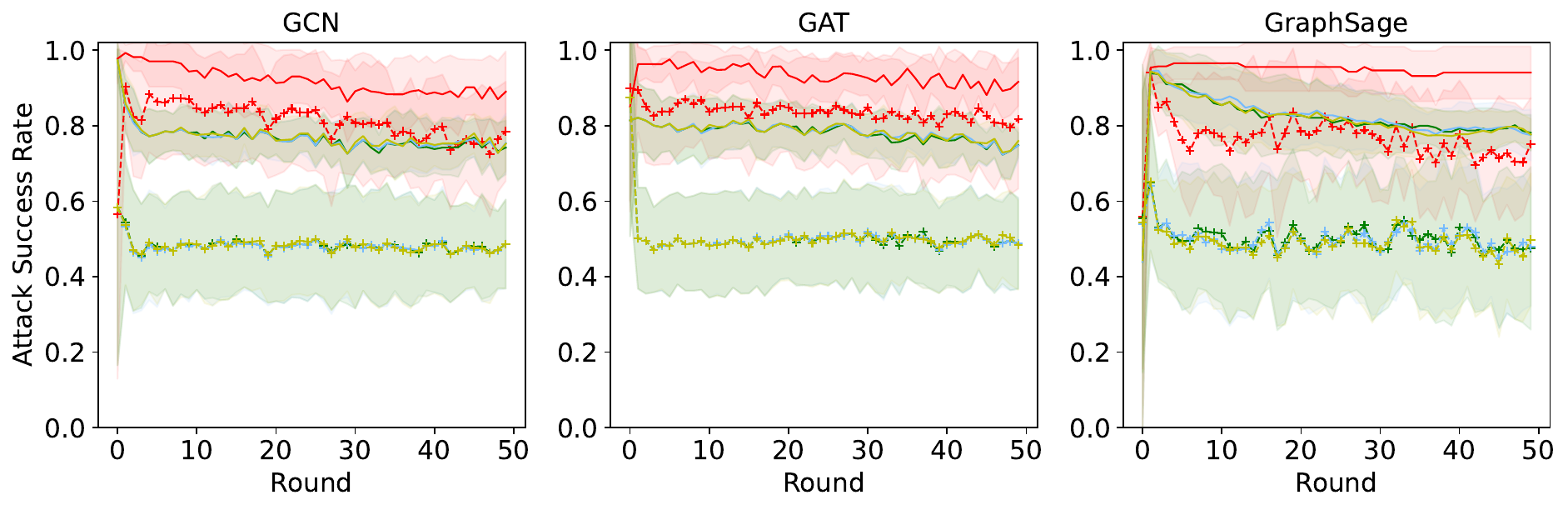}
         \caption{PROTEINS\_full}
         \label{fig:bk_malicious_majority_b}
     \end{subfigure}
     \hfill
     \begin{subfigure}[b]{0.48\textwidth}
         \centering
         \includegraphics[width=\textwidth]{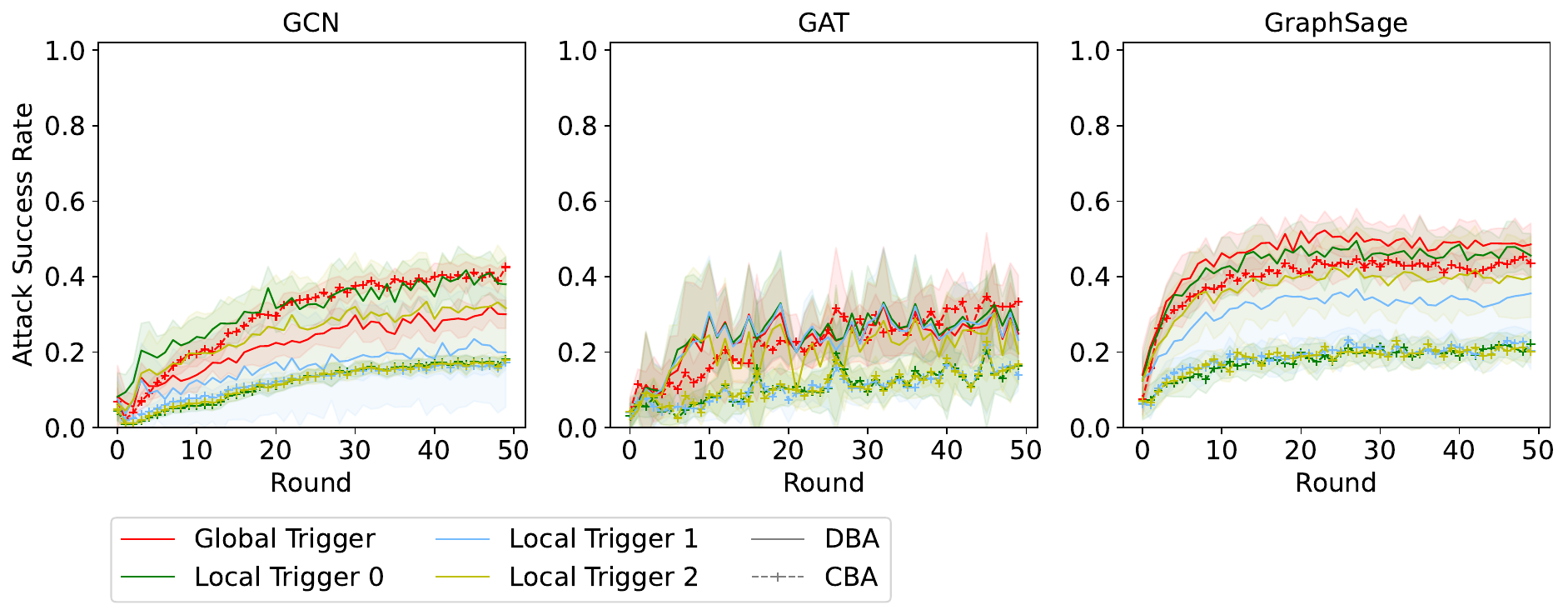}
         \caption{TRIANGLES}
         \label{fig:bk_malicious_majority_c}
     \end{subfigure}
     \caption{Backdoor attack results in the malicious majority attack scenario.}
     \label{fig:bk_malicious_majority}

\end{figure}

\textbf{Malicious Majority Attack Scenario}
\label{exp:malicious_majority}
Figure~\ref{fig:bk_malicious_majority} illustrates the attack results in the malicious majority attack scenario. 
Compared with the honest majority attack scenario, in most cases, the attack success rate of DBA and CBA increases
as with more malicious clients, more malicious updates are uploaded to the global model, making the attack more effective and persistent. 
Moreover, the increase in DBA is more significant than in CBA. For instance, based on the NCI1 dataset and GAT model, the DBA ASR with the global trigger in the honest majority attack scenario is $17.54\%$ higher than CBA, while in the malicious majority attack scenario, the ASR difference is $20.65\%$. Thus, increasing the number of malicious clients is more beneficial for DBA than CBA. With more malicious clients, more local models are used to learn the trigger patterns in DBA, while there is only one malicious local model in CBA.
 
For CBA, the ASR with the global trigger is higher while the attack performance with local triggers stays at a similar level or even decreases. One possible reason is that more malicious clients mean a larger global trigger, requiring more learning capacity of the model. If there is not enough learning capacity for every local trigger in the global trigger, the backdoored model can have poor attack performance with a specific local trigger but will behave well with the union set of the local triggers, i.e., the global trigger.

\textbf{Impact of the Number of Clients}
\label{exp:impact_of_k}
We only set the number of clients as $5$ for these graph datasets because some of these datasets, i.e., NCI1 and PROTEINS\_full, are small (less than $5,000$ graphs). However, to explore the impact of the number of clients on DBA and CBA, we also conduct experiments with more clients on the largest dataset - TRIANGLES.
We set the number of clients as $10$ and $20$ and keep the ratio of malicious clients among the total clients the same as before, i.e., $0.4$ and $0.6$ for the honest majority and malicious majority attack scenarios, respectively. Here, we provide the results of the honest majority attack scenario, as shown in Figure~\ref{fig:more_clients_honest_majority}. The results of the malicious majority attack scenario are given in Appendix~\ref{sec:appendix_more_clients}, and the phenomenon between the two attack scenarios with $10$ and $20$ clients is similar to that with $5$ clients.

\begin{figure}[htbp]
     \centering
     \begin{subfigure}[b]{0.48\textwidth}
         \centering
         \includegraphics[width=\textwidth]{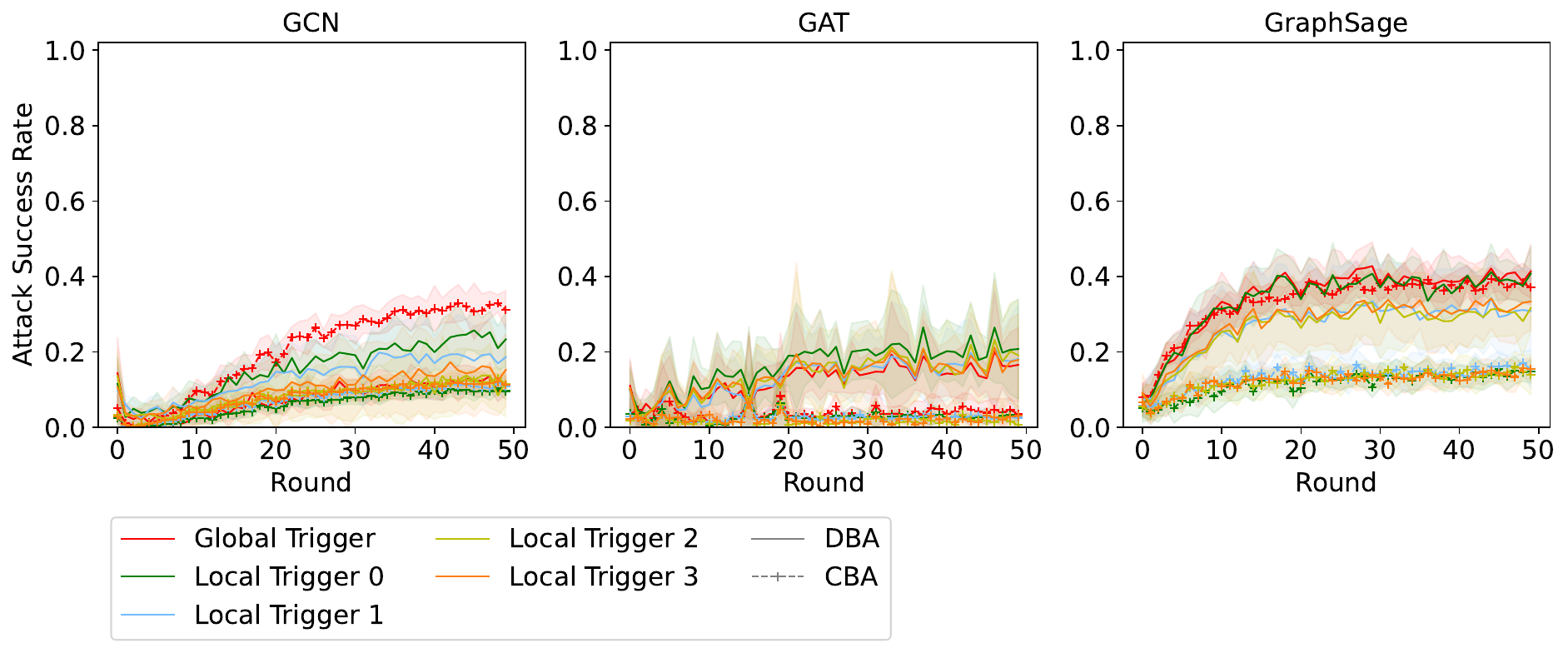}
         \caption{10 clients}
         \label{fig:more_clients_honest_majority_a}
     \end{subfigure}
     \begin{subfigure}[b]{0.48\textwidth}
         \centering
         \includegraphics[width=\textwidth]{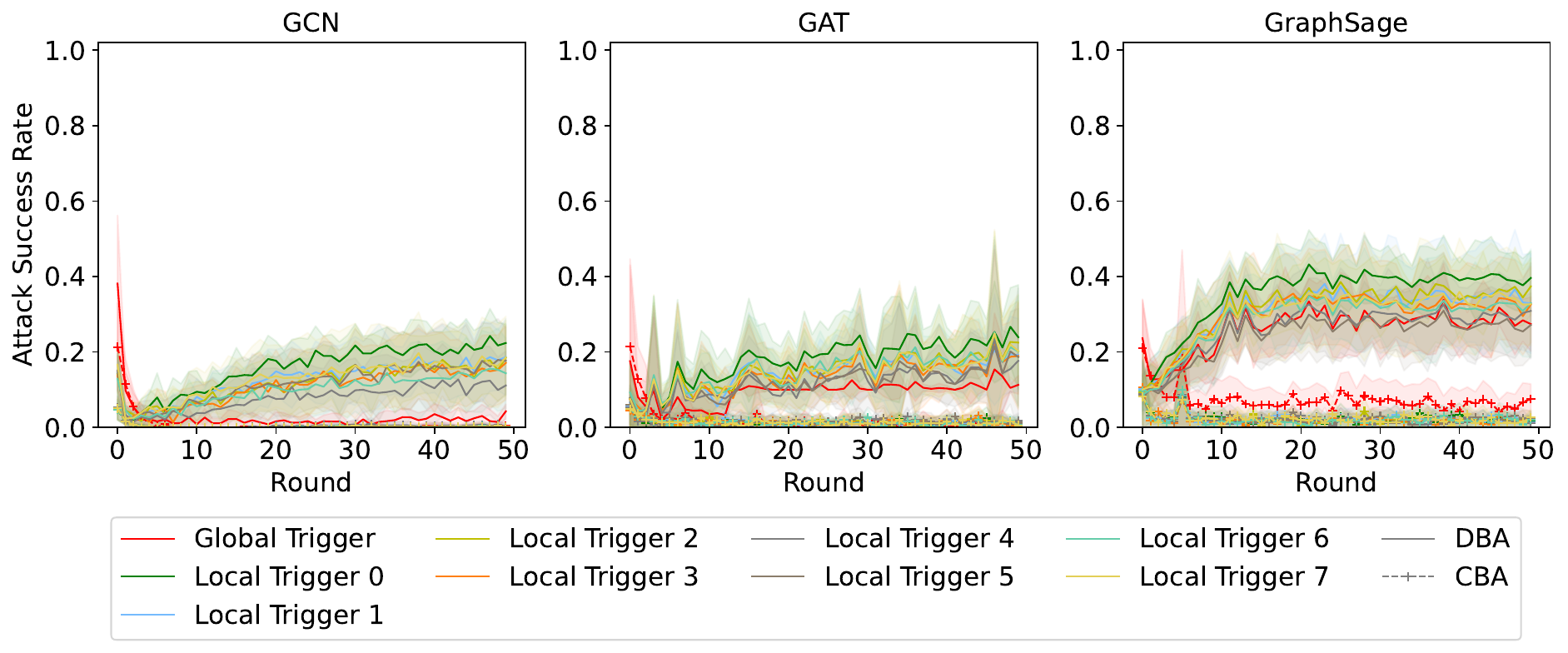}
         \caption{20 clients}
         \label{fig:more_clients_honest_majority_b}
     \end{subfigure}
     \caption{Backdoor attack results of TRIANGLES with more clients in the honest majority attack scenario.}
     \label{fig:more_clients_honest_majority}
\end{figure}

It is obvious that with the increase in the number of clients, the attack success rate of CBA decreases dramatically while the attack performance of DBA keeps steady. This is reasonable because, in CBA, there is only one malicious client whose malicious updates contribute less to the global model with more clients in total. On the contrary, in DBA, the proportion of malicious clients among total clients is the same, meaning the malicious updates contribute the same to the global model regarding the different number of clients. Therefore, as shown in Figures~\ref{fig:more_clients_honest_majority_a} and~\ref{fig:more_clients_honest_majority_b}, the number of clients has negligible impact to the DBA.

\textbf{Impact of the Percentage of Malicious Clients}
\label{exp:impact_of_m}
Although we have analyzed the experiments with the honest majority and malicious majority scenarios, we further explore the impact of the percentage of malicious clients on the attack performance by calculating their Pearson Correlation Coefficient (PCC), as shown in Figure~\ref{fig:NM} in Appendix~\ref{sec:appendix-impact-asr-m}, (we provide the results for the GraphSage model as the example as they are more stable, and the results of other models are aligned). Recall that $M$ represents the number of malicious clients, and each number over the line is the corresponding PCC. As we can see, for all datasets, PCC in DBA is larger than CBA, meaning the increase in $M$ has a more positive impact on DBA than CBA. 
This is intuitive as more malicious clients in DBA lead to more local models embedded with local triggers, while in CBA, it means a larger global trigger due to more local triggers.
Specifically, in DBA, more malicious clients mean more model weights to learn the trigger. In CBA, there is only one attacker, and learning a larger global trigger can be out of the model's representation capability.
Additionally, as we keep the poisoning intensity of DBA and CBA the same for each malicious client, there are more poisoned training data in DBA than in CBA as more malicious clients are used.

\begin{figure*}[htbp]
    \centering
    \includegraphics[width=0.88\textwidth]{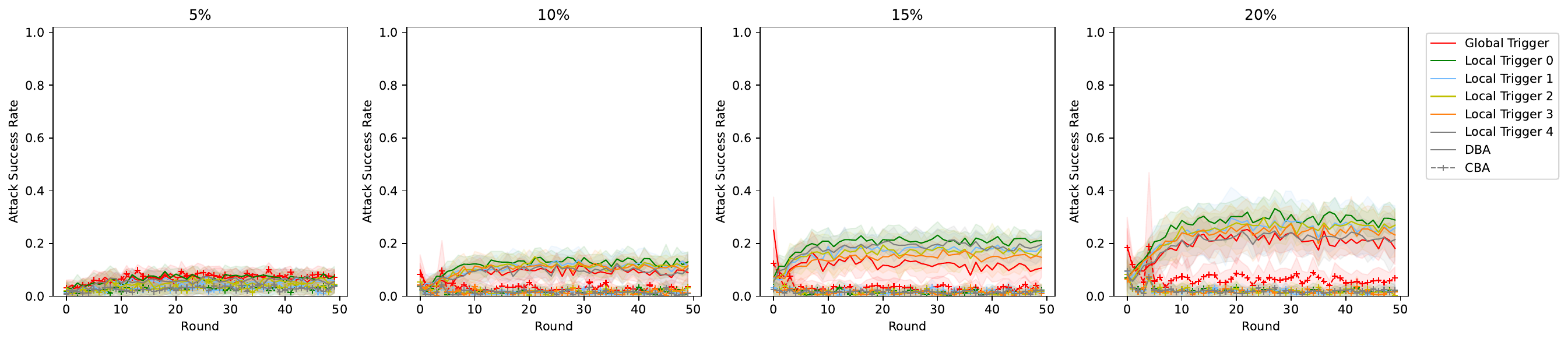}
    \caption{Backdoor attack results of TRIANGLES with less percentage of malicious clients ($K=100$, GraphSage).}
    \label{fig:less_percentage_nm}
\end{figure*}

We also explore the attack performance with more clients and less percentage of malicious clients on the large dataset - TRIANGLES.
Figure~\ref{fig:less_percentage_nm} shows the attack results on TRIANGLES with $100$ clients and fewer malicious clients, ranging from $5\%$ to $20\%$ (here, we also take the results of the GraphSage model as the example, the results of other models are presented in Appendix~\ref{sec:less_malicious_clients}). Table~\ref{table:asr_less_nm} illustrates the specific attack results.
We can see from Figure~\ref{fig:less_percentage_nm} that DBA's ASR gradually increases with more malicious clients while CBA's ASR stays below $10\%$, further verifying that the increase in $M$ has a more positive impact on DBA than CBA. 
From Figure~\ref{fig:less_percentage_nm}, we can still observe that with $20\%$ malicious clients, DBA can also achieve ASR of more than $20\%$, which means with less percentage (e.g., $20\%$) of malicious clients, the DBA is still effective. 
With more clients in total, the attack performance of CBA decreases, consistent with the observation in Figure~\ref{fig:more_clients_honest_majority}.
Thus, adding more clients does not change our previous conclusions (with $5$ clients).

\begin{table}[!htb]
\footnotesize
 \centering
 \caption{Attack success rate of CBA and DBA with less percentage of malicious clients in TRIANGLES ($K$=100, GraphSage).}
 \resizebox{.45\textwidth}{!}{%
\begin{tabular}{c|c|c|c|c} 
 \hline
\multirow{2}{*}{Model} & \multicolumn{4}{c}{Attack Success Rate (CBA\% | DBA\%)} \\
 \cline{2-5}
 & $5\%$ & $10\%$ & $15\%$ & $20\%$ \\
 \hline
 \hline
 GCN & $6.50$ | $2.03$ & $1.36$ | $1.39$ & $1.28$ | $1.06$ & $1.52$ | $1.95$\\
 \hline
 GAT & $1.48$ | $3.91$ & $1.11$ | $3.65$ & $1.22$ | $6.97$ & $1.51$ | $8.01$ \\
 \hline
 GraphSage & $7.64$ | $6.89$ & $2.85$ | $8.80$ & $2.87$ | $10.93$ & $5.53$ | $20.49$ \\
 \hline
 \end{tabular}
}
\label{table:asr_less_nm}
\end{table}

\textbf{Analysis of CBA results}
In Figure~\ref{fig:bk_honest_majority}, for CBA, the attack success rate of all local triggers can be as high as the global trigger, which is counterintuitive as the centralized attack only embeds the global trigger into the model.
To explain these results, we further implement an experiment (NCI1 on GraphSage model) where we evaluate the attack success rate of the global trigger and local triggers in both the malicious local model~\footnote{For the CBA, we assume there is one centralized attacker, so there is only one local model that will be poisoned and we define this model as the malicious local model} and the global model. 
As shown in Figure~\ref{fig:CBA_analysis}, in the malicious local model, the ASR of all local triggers is already close to the global trigger, which means that the malicious local model has learned the pattern of each local trigger.
After aggregation, the global model inherits the capacity of local models. Once any local trigger exists, the global model will misclassify the data sample into the attacker-chosen target label.

Still, in~\cite{xie2019dba}, for the CBA, the attack success rate of all local triggers is significantly lower than the global trigger.
There, the malicious local model learns the global trigger instead of each local trigger, so the poisoned model can only misclassify the data sample once there is a global trigger in the data.
The different results in CBA between~\cite{xie2019dba} and our work can be explained since there, the local triggers composing the global trigger are located close to each other (i.e., less than three pixels distance). In our work, the location of local triggers is random since a graph is non-Euclidean data where we cannot put nodes in some order. When the local trigger graphs are further away from each other, the malicious local model in CBA can only learn the local trigger instead of the global trigger. 

\subsection{Clean Accuracy Drop}

The goal of the backdoor attack is to make the backdoored model simultaneously fit the main task and backdoor task. Therefore, it is critical that the trained model still behaves normally on untampered data samples after training with the poisoned data. Here, we use \textit{clean accuracy drop (CAD)} to evaluate if the backdoored model can still fit the original main task.
CAD is the classification accuracy difference between global models with and without malicious clients over the clean testing dataset. 
CBA's and DBA's final clean accuracy drop results in the honest and malicious majority attack scenarios are given in Tables~\ref{table:cad_honest_majority} and~\ref{table:cad_malicious_majority}, respectively. In most cases, both attacks have a low CAD, i.e., around $2\%$, and only in a few cases is there a significant CAD. 
These results imply that, in most cases, both attacks have a negligible impact on the original task of the model. Additionally, in some cases, DBA's CAD is significantly higher than CBA's, e.g., DBA's CAD is $5.74\%$ in the GraphSage model on TRIANGLES while CBA's is $3.24\%$, as shown in Table~\ref{table:cad_malicious_majority}. 
At the same poisoning intensity for each client, there are more poisoned data in DBA than in CBA, leading to worse performance in the main task.
The substantial clean accuracy drop in DBA can also be observed in~\cite{xie2019dba}.


\begin{table}[!htb]
\footnotesize
 \centering
 \caption{Clean accuracy drop of CBA and DBA in the honest majority attack scenario.}
\begin{tabular}{M{60pt}|M{40pt}|M{40pt}|M{40pt}} 
 \hline
 \multirow{2}{*}{Dataset} & \multicolumn{3}{c}{Clean Accuracy Drop (CBA\% | DBA\%)}  \\
 \cline{2-4}
  & GCN & GAT & GraphSage \\
 \hline
 \hline
 NCI1 & $2.54$ | $2.01$ & $0.84$ | $1.42$ & $0.93$ | $0.16$ \\
 \hline
 PROTEINS\_full & $1.81$ | $4.06$ & $0.49$ | $0.46$ & $2.31$ | $2.82$ \\
 \hline
 TRIANGLES & $0.01$ | $1.32$ & $3.71$ | $2.87$ & $3.31$ | $4.45$ \\
 \hline
\end{tabular}
\label{table:cad_honest_majority}
\end{table}

\begin{table}[!htb]
\footnotesize
 \centering
 \caption{Clean accuracy drop of CBA and DBA in the malicious majority attack scenario.}
\begin{tabular}{M{60pt}|M{40pt}|M{40pt}|M{40pt}} 
\hline
 \multirow{2}{*}{Dataset} & \multicolumn{3}{c}{Clean Accuracy Drop (CBA\% | DBA\%)}  \\
 \cline{2-4}
  & GCN & GAT & GraphSage \\
 \hline
 \hline
 NCI1 & $4.45$ | $2.74$ & $1.03$ | $1.07$ & $1.29$ | $2.22$\\
 \hline
 PROTEINS\_full & $2.78$ | $1.30$ & $0.03$ | $2.65$ & $2.72$ | $4.59$ \\
 \hline
 TRIANGLES & $0.14$ | $0.30$ & $5.10$ | $5.84$ & $3.24$ | $5.74$ \\
 \hline
\end{tabular}
\label{table:cad_malicious_majority}
\end{table}

\section{Defense}
\label{sec:defense}

\noindent \textbf{Potential Countermeasures}
FoolsGold~\cite{fung2018mitigating} is a robust FL aggregation algorithm that can identify attackers in federated learning based on the diversity of client updates. It reduces the aggregation weights of detected malicious clients while retaining the weights of other clients. One of the assumptions in this defense is that each client's training data is non-i.i.d and has a unique distribution, which fits the non-i.i.d data distribution setting in our paper. 
Thus, in this work, we focus on evaluating the attack effectiveness of DBA and CBA against FoolsGold.~\footnote{In the previous version of the paper (published at the ACSAC conference~\cite{more-is-better-acsac}), there was a mistake in our implementation of FLAME indicating FLAME not to work against our attack. That statement was incorrect, and this was communicated to the authors of FLAME.} 

\begin{figure}
     \centering
     \begin{subfigure}[b]{0.48\textwidth}
         \centering
         \includegraphics[width=\textwidth]{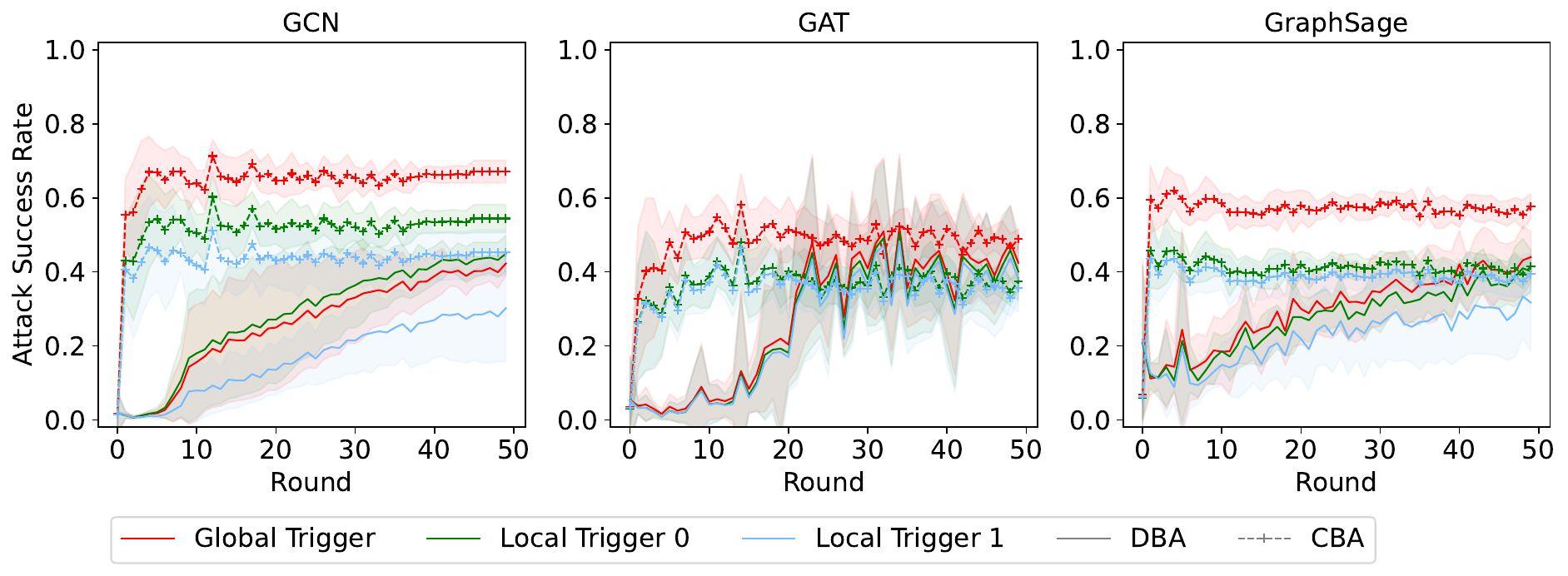}
         \caption{Attack success rate}
         \label{fig:bk_triangles_defense_asr}
     \end{subfigure}
     \hfill
     \begin{subfigure}[b]{0.48\textwidth}
         \centering
         \includegraphics[width=\textwidth]{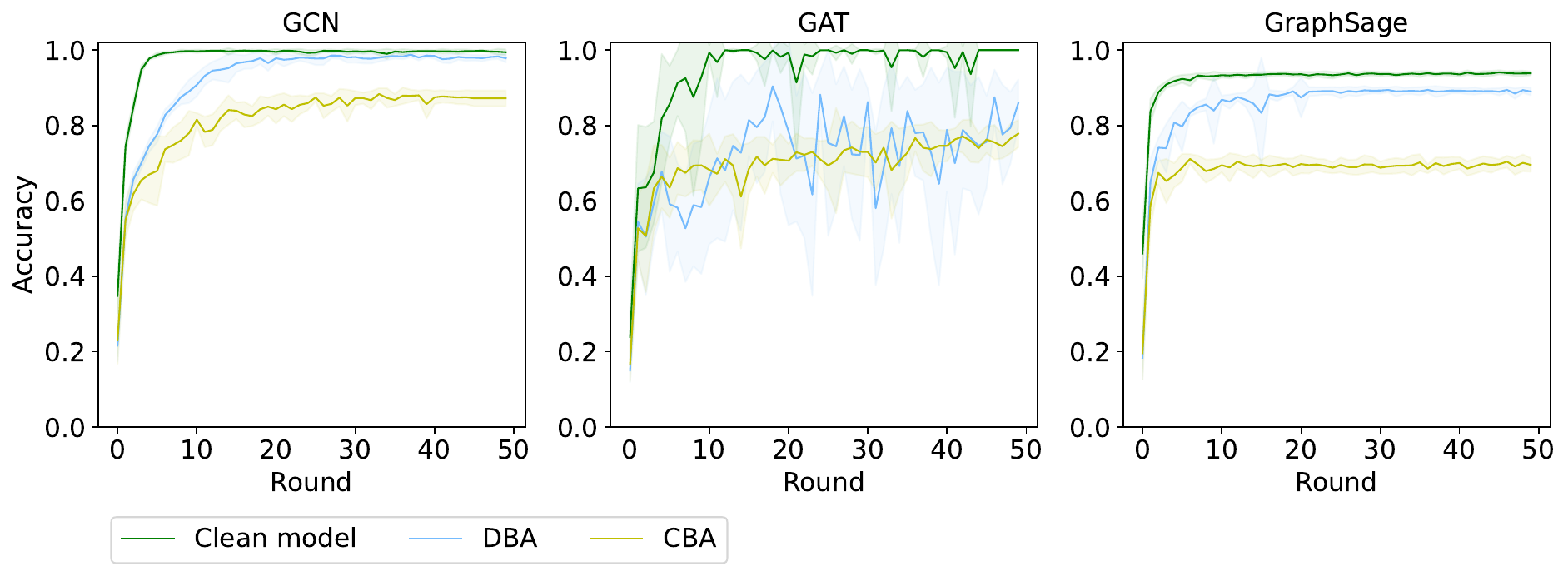}
         \caption{Testing accuracy}
         \label{fig:bk_triangles_defense_test}
     \end{subfigure}     
     \caption{Backdoor attack results of TRIANGLES on FoolsGold for the honest majority.}
     \label{fig:bk_triangles_defense}
\end{figure}

\noindent \textbf{Results and Analysis}
Figure~\ref{fig:bk_triangles_defense} shows the attack performance for the TRIANGLES dataset under FoolsGold in the honest majority attack scenario (the results in the malicious majority attack scenario are similar). The results for other datasets illustrate that FoolsGold has a negligible impact on the attack performance, as shown in Appendix~\ref{sec:appendix-B}. As illustrated in Figure~\ref{fig:bk_triangles_defense_asr}, we can observe that for DBA, the ASR under FoolsGold remains consistent for the GraphSage model, whereas it rises along with a decrease in testing accuracy for the GAT model. 
Moreover, generally, under FoolsGold, there is a significant increase in CBA's ASR in all models, but the testing accuracy of CBA reduces significantly at the same time. 
For example, the CBA's ASR increases by about $20\%$ for the GraphSage model. However, the testing accuracy of CBA on GraphSage has a drop of more than $20\%$ (Figure~\ref{fig:bk_triangles_defense_test}). 
Our hypothesis for this situation is that under FoolsGold, the malicious client in CBA is assigned a higher weight (recall the description of the FoolsGold mechanism from the paragraph above)
than other clients, so malicious updates contribute more to the aggregated model. Simultaneously, the low weights on the honest clients' updates lead to the failure of the performance on the original task.
We reported FoolsGold's weights on every client in DBA and CBA in Appendix~\ref{sec:appendix-C} and showed that this hypothesis is reasonable.
One possible reason is that in CBA, there is only one malicious client whose updates are likely to appear dissimilar from those of other honest clients, so FoolsGold cannot identify the malicious updates successfully.

Based on the experimental results against FoolsGold, we find that the tested defense cannot detect malicious updates successfully. 
One reason may be that this defense applies \textit{cosine distance} to try to identify malicious models, i.e., the distance between malicious updates is smaller between honest updates. Still, in our attacks, the malicious clients' updates could already be very dissimilar to each other, so the malicious updates are likely to be clustered into honest updates.
%


\section{Related Work}
\label{sec:related}

\noindent \textbf{Backdoor Attacks in GNNs}
Several recent works have conducted backdoor attacks on GNNs. Zhang et al. proposed a subgraph-based backdoor attack on GNNs for the graph classification task~\cite{DBLP:conf/sacmat/ZhangJWG21}. 
Xi et al. presented a subgraph-based backdoor attack on GNNs, that works for both node classification and graph classification tasks~\cite{DBLP:conf/uss/XiPJ021}. 
Xu et al. investigated the explainability of the impact of the trigger injecting position on the performance of backdoor attacks on GNNs and proposed a new backdoor attack strategy for the node classification task~\cite{DBLP:conf/wisec/XuXP21}. 
\textit{All current attacks are implemented in centralized training for GNNs. No works explore the backdoor attacks in distributed training for GNNs, e.g., Federated GNNs.}

\noindent 
\textbf{FL on GNNs}
FL has gained increasing attention as a training paradigm where data is distributed at remote devices and models are collaboratively trained in a central server. While FL has been widely studied in Euclidean data, e.g., images, texts, and sound, there are increasing studies about FL in graph data. FL on graph data was introduced in~\cite{DBLP:journals/corr/abs-1901-11173}, where each client is regarded as a node in a graph.
When it comes to detecting financial crimes (e.g., fraud or money laundering), traditional machine learning tends to lead to severe overreporting of suspicious activities. Thanks to the reasoning ability of the graph neural network, its advantages can be well-reflected. 
Considering the need for privacy,~\cite{suzumura2019towards} proposed the framework for Federated GNNs to optimize the machine learning model.
Besides, other research works~\cite{jiang2020federated, zhou2021vertically, wu2021fedgnn} have been dedicated to enhancing the security of Federated GNNs. By using secure aggregation,~\cite{jiang2020federated} proposed a method to predict the trajectories of objects via aggregating both spatial and dynamic information without information leakage. With differential privacy,~\cite{zhou2021vertically} and~\cite{wu2021fedgnn} put forward a framework to train Federated GNNs for vertical FL and recommendation system, respectively. 
Moreover, SpreadGNN was proposed in~\cite{he2021spreadgnn} to perform FL without a server. 
\textit{Although there is an increasing number of works on FL for graph data, the vulnerability of Federated GNNs to backdoor attacks is still underexplored}.  

\noindent 
\textbf{The Security Assumption of Malicious Majority Clients}
Cao et al. took into account the situation of backdoor attacks in the malicious majority of clients and proposed a  method of defense-\emph{FLTrust}~\cite{cao2021fltrust}. 
Before training begins, an honest server collects and trains on a small dataset. The server takes the updates obtained by training on a small dataset as the root of trust in each iteration. It is then compared to the updates uploaded by the clients. If the cosine similarity between them is too small, the updates will be filtered out.
With this approach, the accuracy of the global model remains equivalent to that of the baseline. 
Based on \emph{FLTrust}, Dong et al. considered the setting of two semi-honest servers and malicious majority clients and proposed \emph{FLOD} to ensure that gradients are not leaked on the server side~\cite{10.1007/978-3-030-88418-5_24}.  
\section{Conclusions and Future Work}
\label{sec:conclusions}

This paper explores how Centralized and Distributed Backdoor attacks behave in Federated GNNs. 
Through extensive experiments on three datasets and three popular GNN models, we showed that generally, DBA achieves a higher attack success rate than CBA. 
We showed that in CBA, the ASR of local triggers could be as high as the global trigger even if, during training, only the global trigger is embedded in the model.
The impact of the percentage of malicious clients on DBA's ASR is analyzed with correlation, where we confirm the intuition that more malicious clients lead to more successful attacks. 
We analyzed the critical backdoor hyperparameters to explore their impact on the attack performance and the main task.
We also demonstrated that DBA and CBA are robust against one defense for the backdoor attack in FL. Interestingly, the CBA's ASR is even higher under the defense. 
The experimental setting in this work verifies the effectiveness of our method in a cross-silo federated learning setting and motivates further research in exploring backdoor attacks in Federated GNNs considering cross-device FL~\cite{shejwalkar2022back}.
Future work will include exploring backdoor attacks in Federated GNNs for the node classification task. For example, in a social media app where each user has a local social network $G^k$ and $\{ G^k \}$ constitutes the latent entire human social network $G$, the developers can train a fraud detection GNN model through FL. In such a case, an attacker can conduct a backdoor attack to force the trained global model to classify a fraud node as benign. 


\begin{acks}
This work was partly supported by the China Scholarship Council and the  European Union's Horizon 2020 Research and Innovation Programme under Grant No. 101021727.
\end{acks}

\bibliographystyle{ACM-Reference-Format}
\bibliography{sample.bib}

\appendix
\section{Notation}
\label{sec:appendix-notation}

In Table~\ref{tab:notations}, we summarize the notations used throughout the paper.

\begin{table}[H]
 \caption{Notations used in this paper.}
 \label{tab:notations}
 \resizebox{.45\textwidth}{!}{%
 \begin{tabular}{c|l} 
 \hline
 \textbf{Notations} & \textbf{Descriptions} \\ \hline
 $y_t$ & target label \\ \hline
 $G_{t}$ & joint global model at round $t$  \\ \hline
 $E$ & local epochs \\ \hline
 $K$ & number of clients \\ \hline
 $M$ & number of malicious clients \\ \hline
 $C_h, C_m$ & honest clients, malicious clients \\ \hline
 $D_{local}$ & client's local training dataset splitted from dataset $D_{train}$ \\ \hline
 $D_{test}$ & testing dataset splitted from dataset $D$ \\ \hline
 $t_{global}$ & global trigger \\ \hline
 $t_{local}$ & local trigger \\ \hline
 $w_t^k$ & client $k$'s local trained model at round $t$ \\ \hline
 $r$ & poisoning ratio\\ \hline
 $s$ & number of nodes in graph trigger \\ \hline
 $\rho$ & edge existence probability in graph trigger \\ \hline
 $D_{trigger}$ & dataset with trigger embedded \\ \hline
 $D_{clean}$ & clean training dataset \\ \hline
 $D_{backdoor}$ & backdoored training dataset\\ \hline
 $B$ & local minibatch size \\ \hline
 $\eta$ & learning rate \\ \hline
\end{tabular}
}
\end{table}

\section{Dataset Statistics}
\label{sec:appendix-dataset-statistic}
In Table~\ref{table:dataset_statistics}, we show various statistics about the datasets used.

\begin{table*}
\footnotesize
 \centering
 \caption{Datasets statistics.}
 \resizebox{.70\textwidth}{!}{%
 \begin{tabular}{M{80pt}|M{40pt}|M{50pt}|M{50pt}|M{25pt}|M{90pt}} 
 \hline
 Dataset & \# Graphs & Avg. \# nodes & Avg. \# edges & Classes & Class Distribution\\
 \hline
 \hline
 NCI1 & $4,110$ & $29.87$ & $32.30$ & $2$ & $2,053[0], 2,057[1]$ \\
 \hline
 PROTEINS\_full & $1,113$ & $39.06$ & $72.82$ & $2$ & $663[0], 450[1]$ \\
 \hline
 TRIANGLES & $45,000$ & $20.85$ & $32.74$ & $10$ & $4,500[0-9]$\\
 \hline
\end{tabular}
}
\label{table:dataset_statistics}
\end{table*}

\section{Analysis of Backdoor Hyperparameters}
\label{section:analysis-of-factors}

\begin{figure*}[!ht]
        \centering
        \begin{subfigure}[b]{0.45\textwidth}
            \centering
            \includegraphics[width=\textwidth]{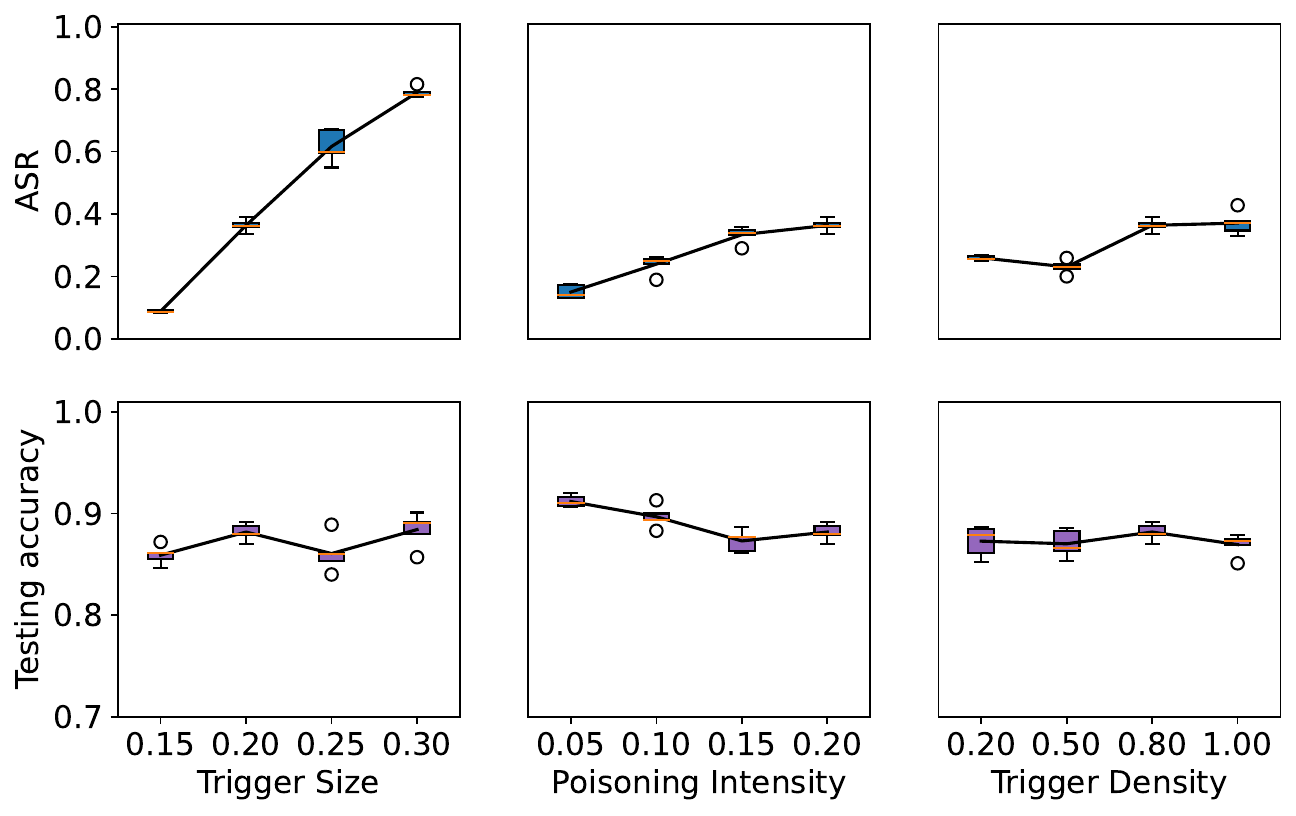}
            \caption{DBA in honest majority attack scenario}
            \label{fig:trigger params14}
        \end{subfigure}
        \begin{subfigure}[b]{0.45\textwidth}  
            \centering 
            \includegraphics[width=\textwidth]{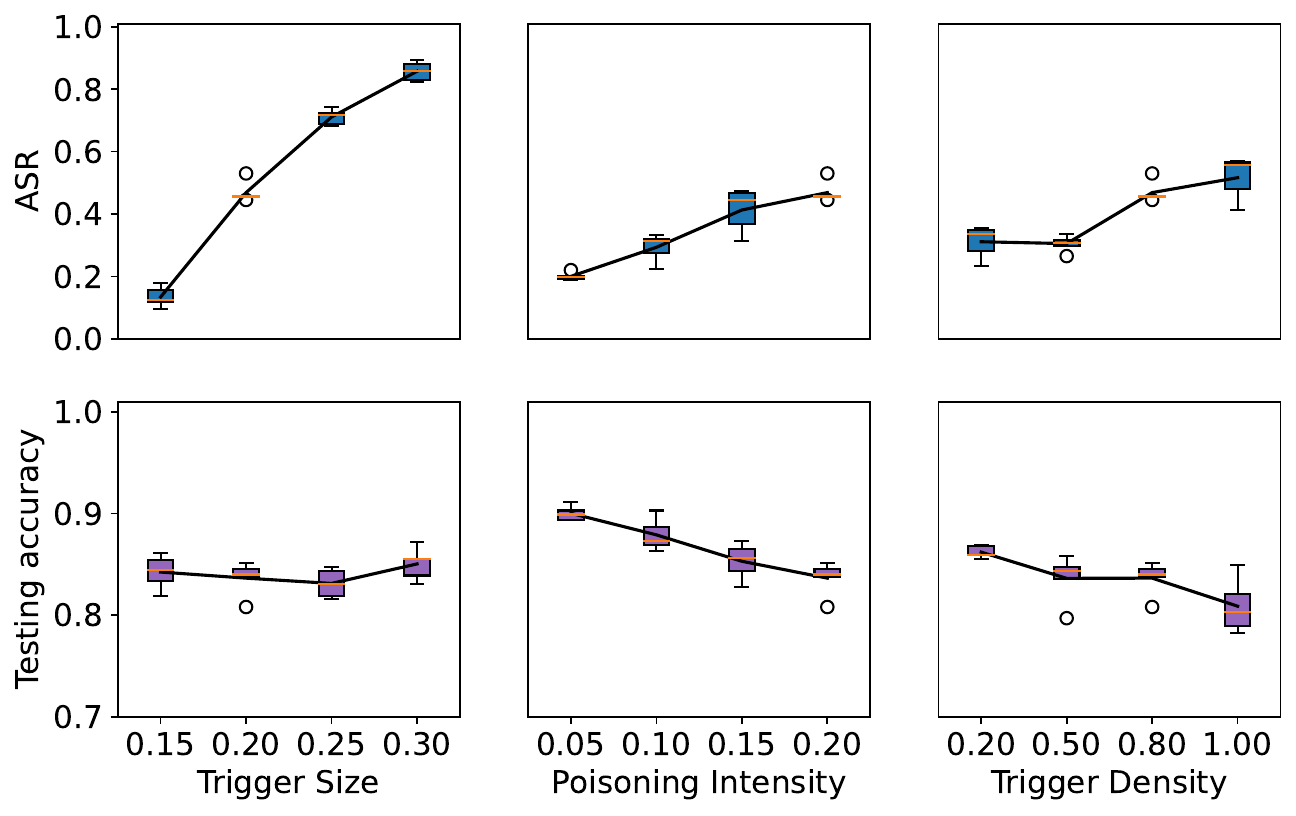}
            \caption{DBA in malicious majority attack scenario}
            \label{fig:trigger params24}
        \end{subfigure}
        \vskip\baselineskip
        \begin{subfigure}[b]{0.45\textwidth}   
            \centering 
            \includegraphics[width=\textwidth]{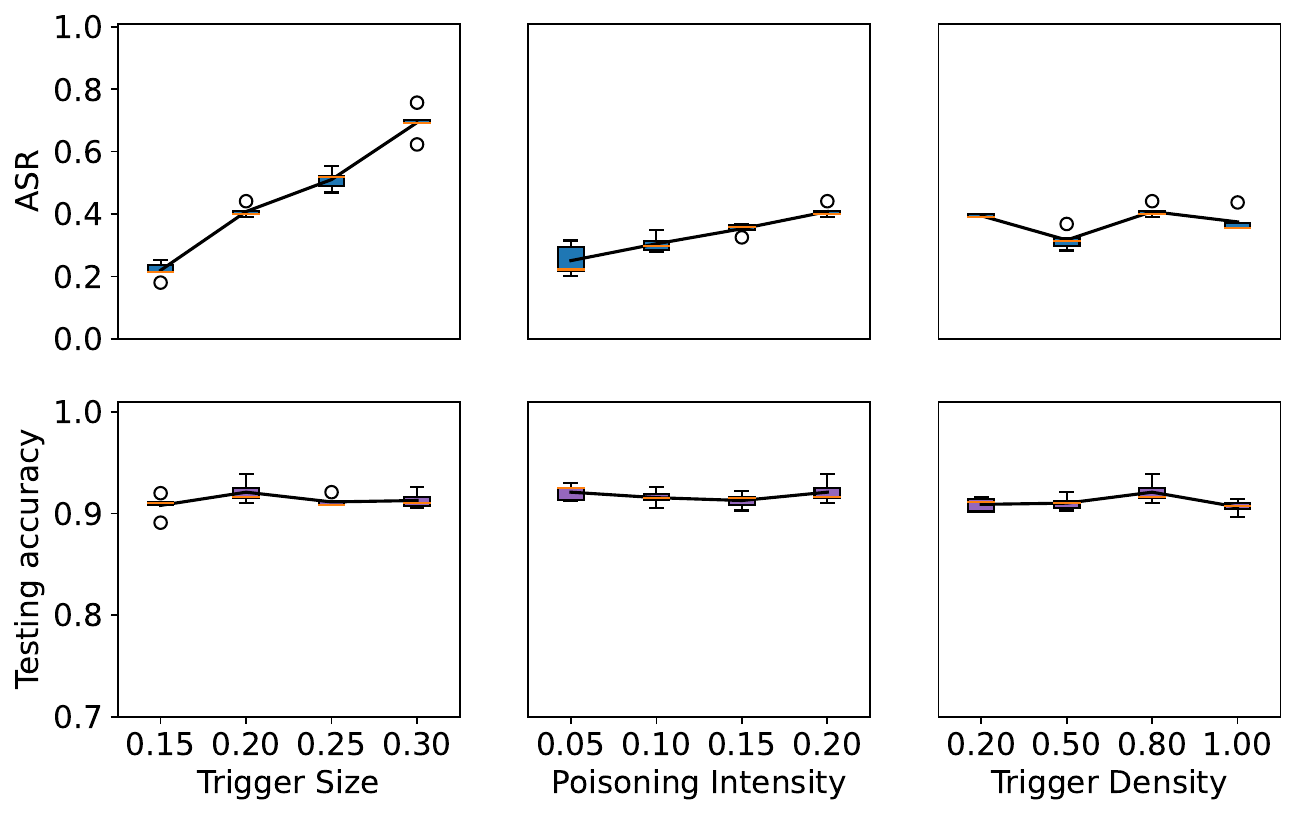}
            \caption{CBA in honest majority attack scenario}
            \label{fig:trigger params34}
        \end{subfigure}
        \begin{subfigure}[b]{0.45\textwidth}   
            \centering 
            \includegraphics[width=\textwidth]{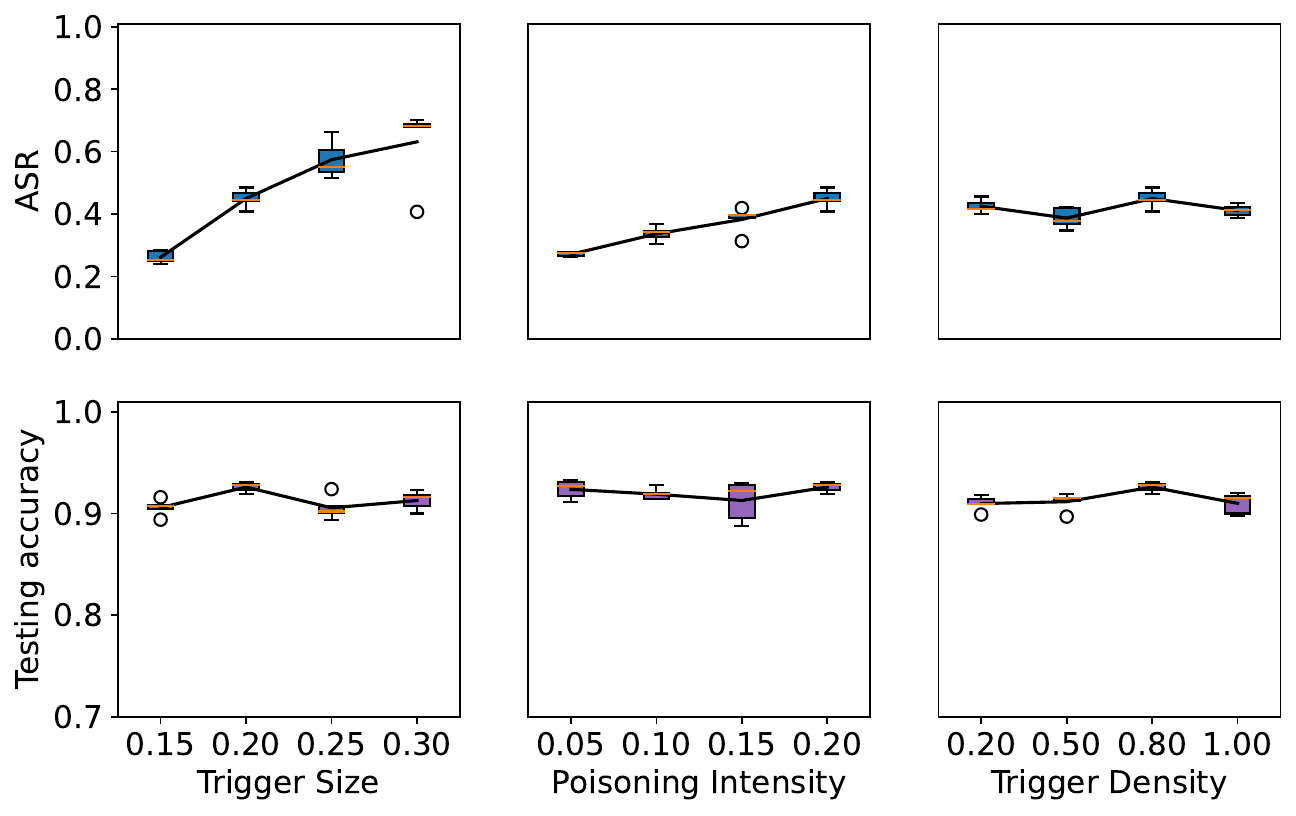}
            \caption{CBA in malicious majority attack scenario}
            \label{fig:trigger params44}
        \end{subfigure}
        \caption{Results on TRIANGLES with different trigger parameters.}
        \label{fig:trigger params}
\end{figure*}

This section studies the backdoor hyperparameters discussed in Section~\ref{subsec:threat}. We only modify one factor for each experiment and keep other factors as in Section~\ref{section:5.1}.
We provide results for TRIANGLES and the GraphSage model as an example as those results are more stable, i.e., have the smallest standard error, and the results of other models are aligned.
For each factor, we evaluate the global trigger's ASR and the test accuracy on the clean test dataset. We illustrate the results on TRIANGLES in two attack scenarios to analyze the effects of each factor for DBA and CBA. The results are shown in Figure~\ref{fig:trigger params}. 

\textbf{Effects of Trigger Size}
From the ASR results in Figure~\ref{fig:trigger params}, for both attacks and attack scenarios, with the increase of trigger size, the attack success rate rises significantly, e.g., the DBA's ASR increases from $0.09$ to $0.80$ with trigger size rising from $0.15$ to $0.30$ (honest majority attack scenario).
There is no significant effect of trigger size on the test accuracy of the global model, implying that the trigger size has little impact on the original main task.

\textbf{Effects of Poisoning Intensity}
Similar to the impact of trigger size on the attack success rate, a higher poisoning intensity gives a higher attack success rate. Intuitively, a backdoor attack can perform better with more poisoned data.
Nevertheless, the increase is less significant than that of different trigger sizes. 
Specifically, in comparison with~\cite{DBLP:conf/uss/XiPJ021}, where there is no obvious difference between the impact of poisoning intensity and trigger size, here, a larger trigger size has a more positive influence on ASR than a larger poisoning intensity. We consider this an interesting observation and plan to investigate it in future work.
Moreover, in DBA, the test accuracy decreases with the increasing poisoning intensity, and with more malicious clients, the drop is more significant, as shown in Figures~\ref{fig:trigger params14} and~\ref{fig:trigger params24}.
This can be explained as with higher poisoning intensity, and more malicious clients, more model weights (including some for the original task) are influenced by the malicious trigger patterns, and the performance on the main task degrades more.
We can also observe that with higher poisoning intensity, there is no obvious drop in the testing accuracy for CBA, as presented in Figures~\ref{fig:trigger params34} and~\ref{fig:trigger params44}. Although more local data is poisoned, the other honest clients (the majority part) still guarantee the performance on the main task. 

\textbf{Effects of Trigger Density}
From Figure~\ref{fig:trigger params24}, DBA's ASR improves from $30.10\%$ to $47.96\%$ when the trigger density increases from $0.50$ to $0.80$. This is because the average complexity of the TRIANGLES dataset is $0.16$~\cite{pudlak1988graph}.
Thus, when the trigger density is set close to this value, the difference between the original graph and the trigger graph is harder to distinguish. 
However, the effect of the trigger density in CBA's ASR is not strong. We see a slight fluctuation as the trigger density increases, but its range is very small to be considered a trend. In CBA, we use only one malicious client, and the weak effect of the trigger density is smoothed by the averaging operation. 

In Figure~\ref{fig:trigger params}, in most cases, there is no significant drop in the test accuracy with an increase in the trigger size and trigger density. On the contrary, in the backdoor attacks in centralized GNNs~\cite{DBLP:conf/wisec/XuXP21}, as trigger size increases, the test accuracy decreases. This can be explained as, in FL, the influence of backdoor functionality on the main task is weakened by the aggregation of local models.

\section{Additional Experimental Results}

\subsection{More Clients (Malicious Majority Attack Scenario)}
\label{sec:appendix_more_clients}

\begin{figure*}[!ht]
     \centering
     \begin{subfigure}[b]{0.48\textwidth}
         \centering
         \includegraphics[width=\textwidth]{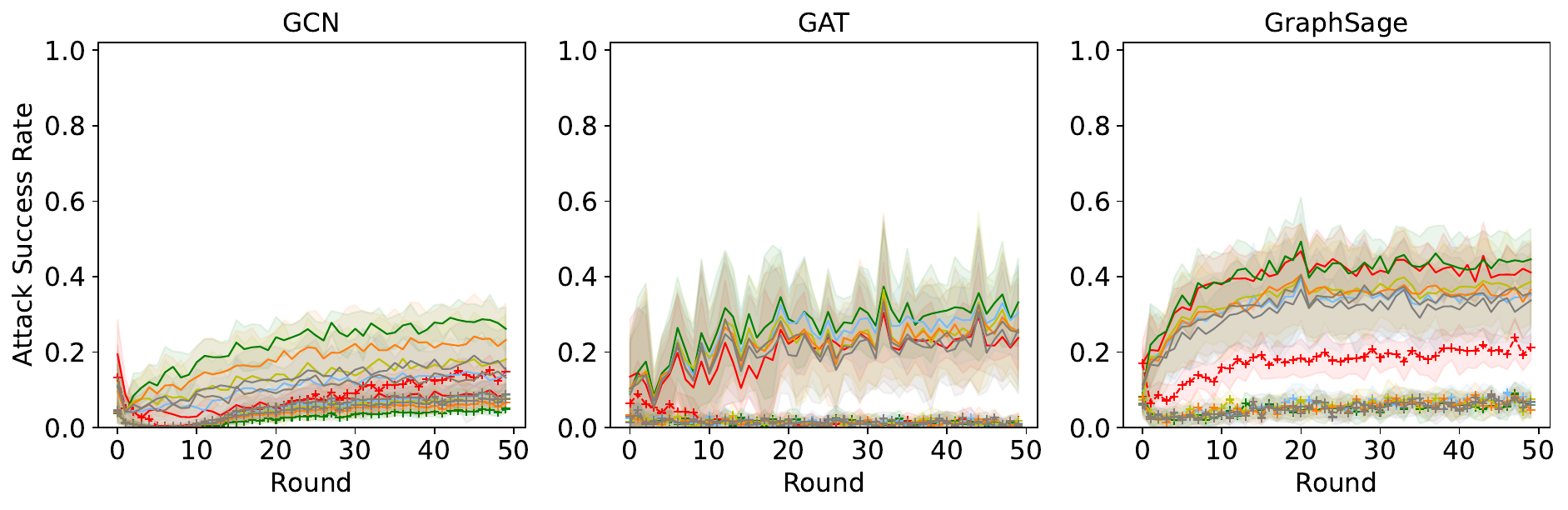}
         \caption{10 clients}
         \label{fig:appx_more_clients_a}
     \end{subfigure}
     \hfill
     \begin{subfigure}[b]{0.48\textwidth}
         \centering
         \includegraphics[width=\textwidth]{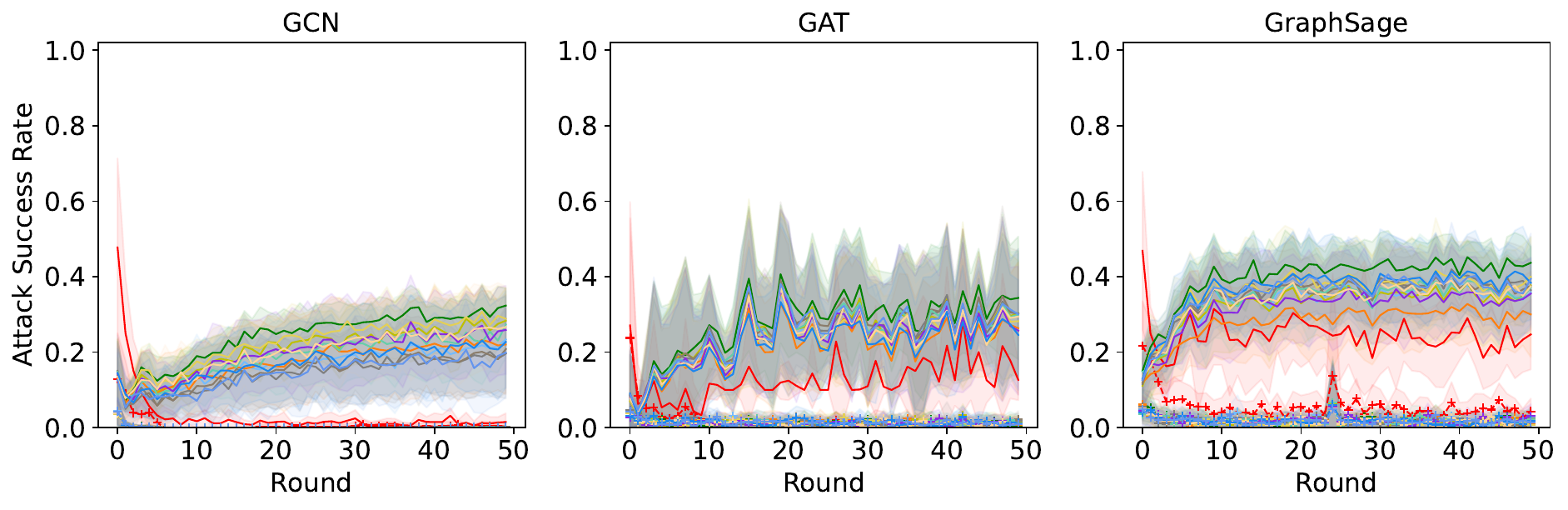}
         \caption{20 clients}
         \label{fig:appx_more_clients_b}
     \end{subfigure}
    \begin{subfigure}[b]{0.48\textwidth}
        \centering
         \includegraphics[width=\textwidth]{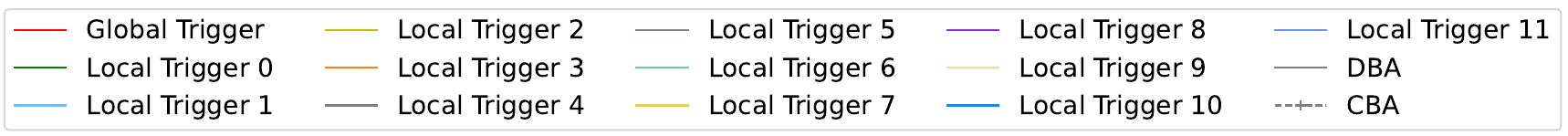}
         \caption{Legend}
         \label{fig:appx_more_clients_c}
     \end{subfigure}
     \caption{Backdoor attack results of TRIANGLES with more clients in the malicious majority attack scenario.}
     \label{fig:appx_more_clients}
\end{figure*}

The attack results on TRIANGLES with $10$ and $20$ clients in the malicious majority attack scenario are shown in Figure~\ref{fig:appx_more_clients}. In the malicious majority attack scenario, with more clients, the ASR of DBA keeps steady while that of CBA drops dramatically, which is consistent with the observations in the honest majority attack scenario, as shown in Figure~\ref{fig:more_clients_honest_majority}.

\subsection{Less Percentage of Malicious Clients}
\label{sec:less_malicious_clients}

\begin{figure*}[htpb]
     \centering
     \begin{subfigure}[b]{0.85\textwidth}
         \centering
         \includegraphics[width=\textwidth]{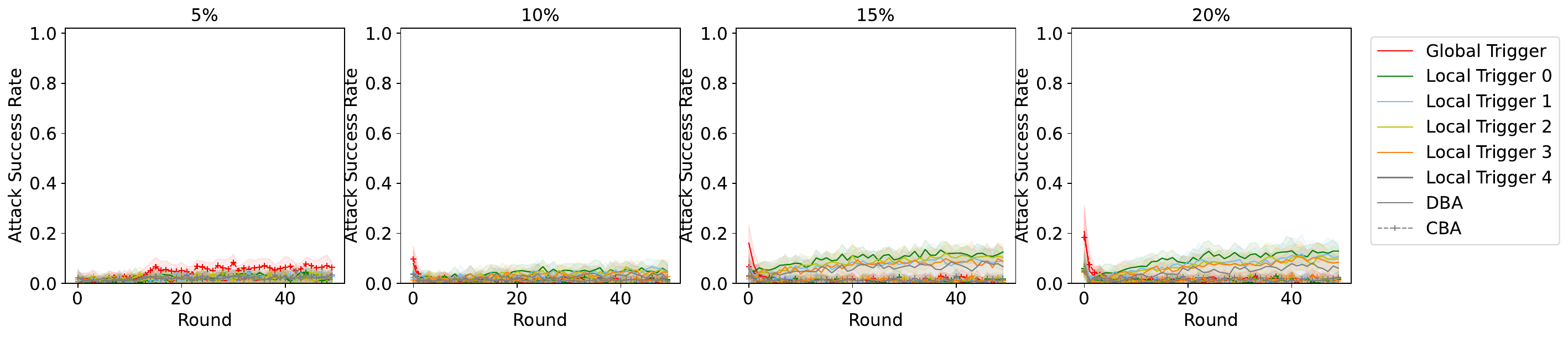}
         \caption{GCN}
         \label{fig:appx_percentage_gcn}
     \end{subfigure}
     \hfill
     \begin{subfigure}[b]{0.85\textwidth}
         \centering
         \includegraphics[width=\textwidth]{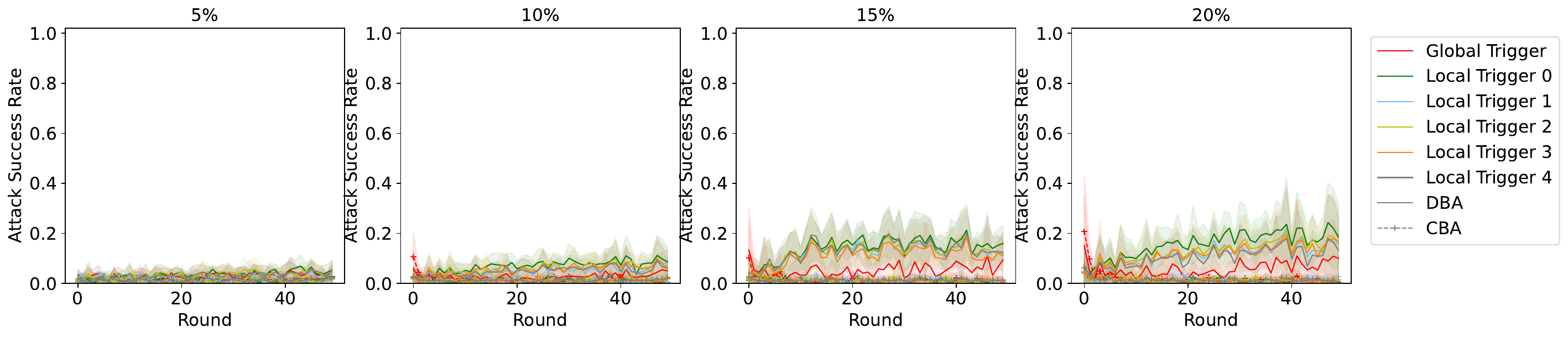}
         \caption{GAT}
         \label{fig:appx_percentage_gat}
     \end{subfigure}
     \caption{Backdoor attack results of TRIANGLES with less percentage of malicious clients ($K=100$, GCN and GAT).}
     \label{Fig:less_nm_gcn_gat}
\end{figure*}

The attack results with less percentage of malicious clients on TRIANGLES are shown in Figure~\ref{Fig:less_nm_gcn_gat}. Similar to the attack results for GraphSage (Figure~\ref{fig:less_percentage_nm})\footnote{Here, we put the first 5 local triggers in the legend to make the figure more clear. The results for the rest local triggers have the same phenomenon}, DBA's ASR is gradually increasing with the rise in the percentage of malicious clients. On the contrary, the attack success rate of CBA keeps steady.

\subsection{Additional Defense Results}
\label{sec:appendix-B}
\begin{figure}[!htpb]
     \centering
     \begin{subfigure}[b]{0.48\textwidth}
         \centering
         \includegraphics[width=\textwidth]{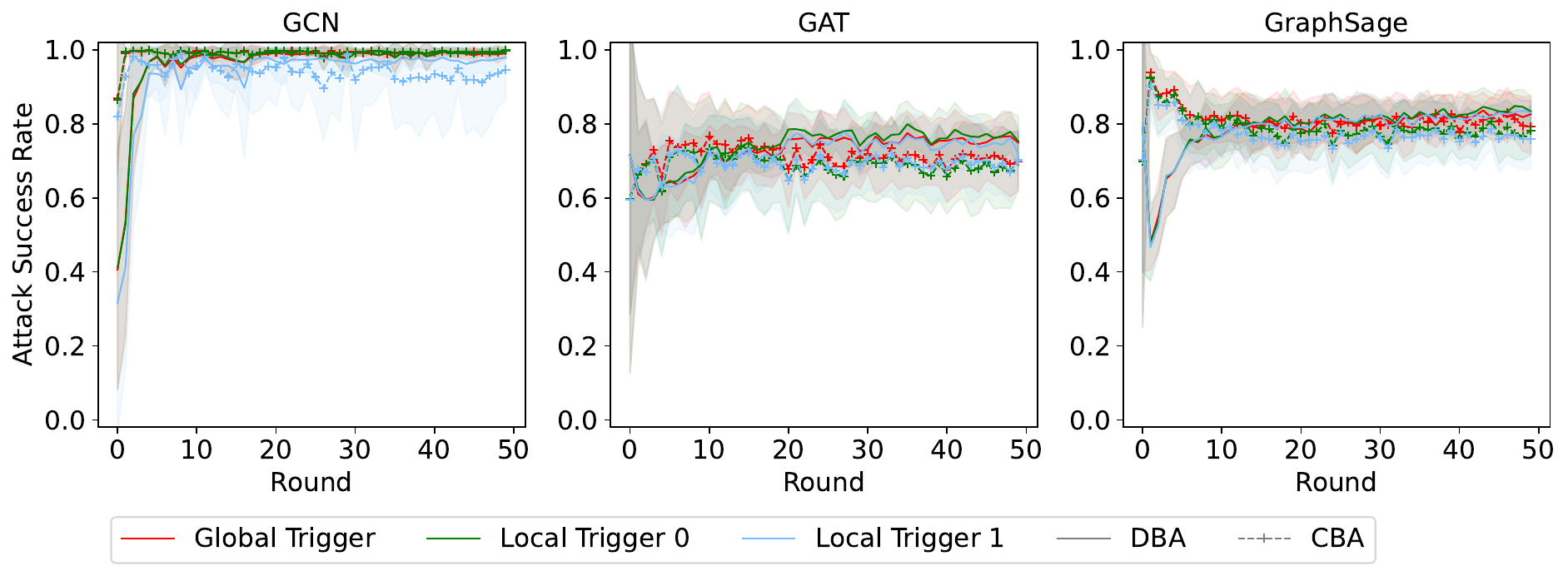}
         \caption{Attack success rate}
         \label{fig:NCI1_defense_asr_honest_majority_a}
     \end{subfigure}
     \hfill
     \begin{subfigure}[b]{0.48\textwidth}
         \centering
         \includegraphics[width=\textwidth]{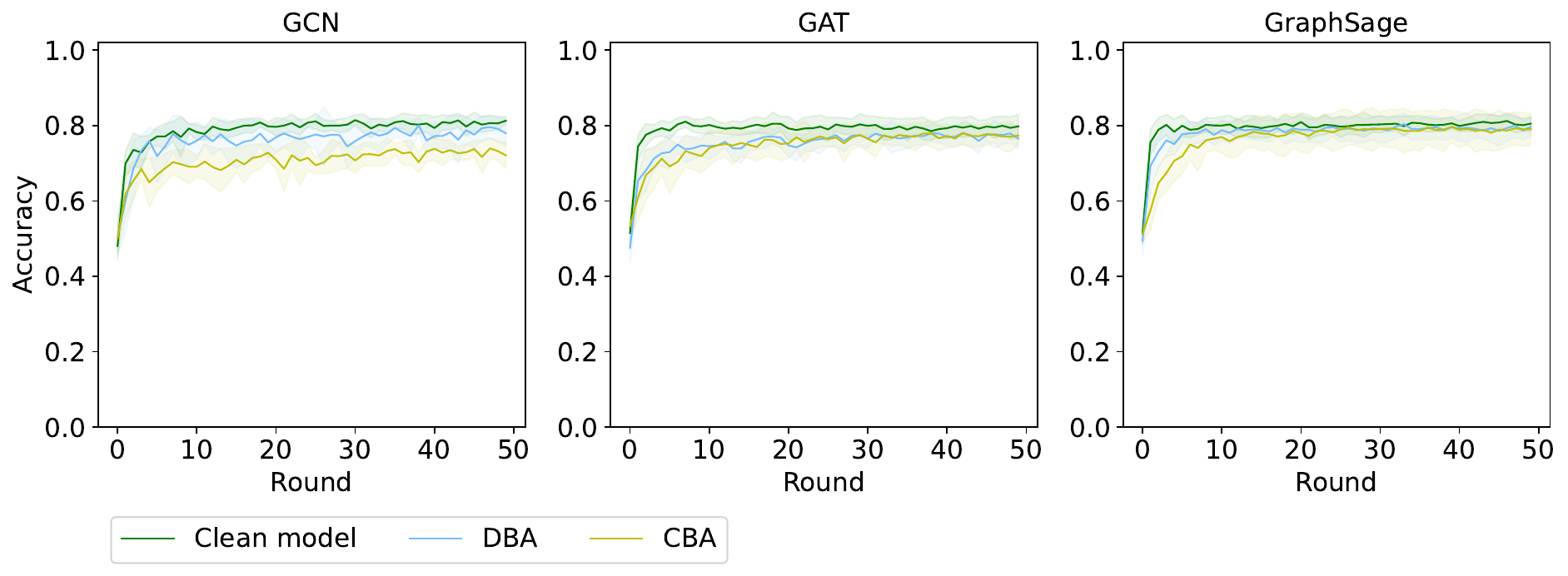}
         \caption{Testing accuracy}
         \label{fig:NCI1_defense_asr_honest_majority_b}
     \end{subfigure}
     \caption{Attack success rate on NCI1 on FoolsGold (in the honest majority attack scenario).}
     \label{fig:NCI1_defense_asr_honest_majority}
\end{figure}

\begin{figure}[!htpb]
     \centering
     \begin{subfigure}[b]{0.48\textwidth}
         \centering
         \includegraphics[width=\textwidth]{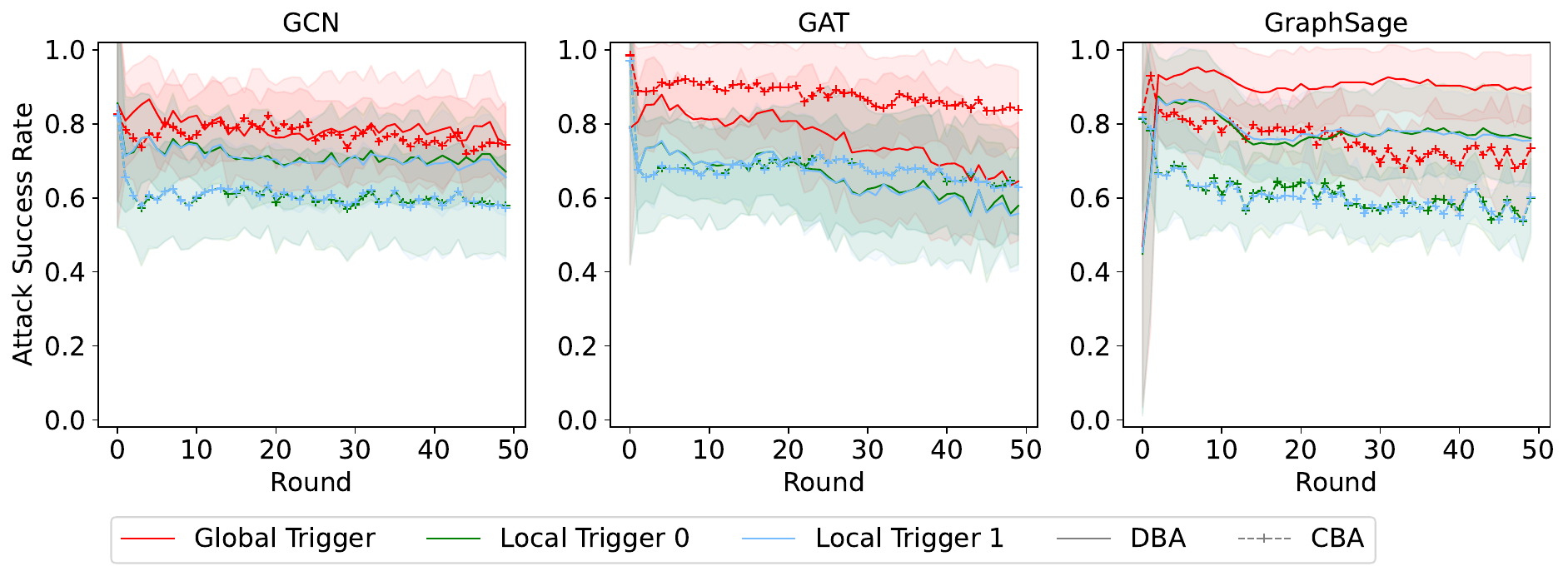}
         \caption{Attack success rate}
         \label{fig:proteins_defense_asr_honest_majority_a}
     \end{subfigure}
     \hfill
     \begin{subfigure}[b]{0.48\textwidth}
         \centering
         \includegraphics[width=\textwidth]{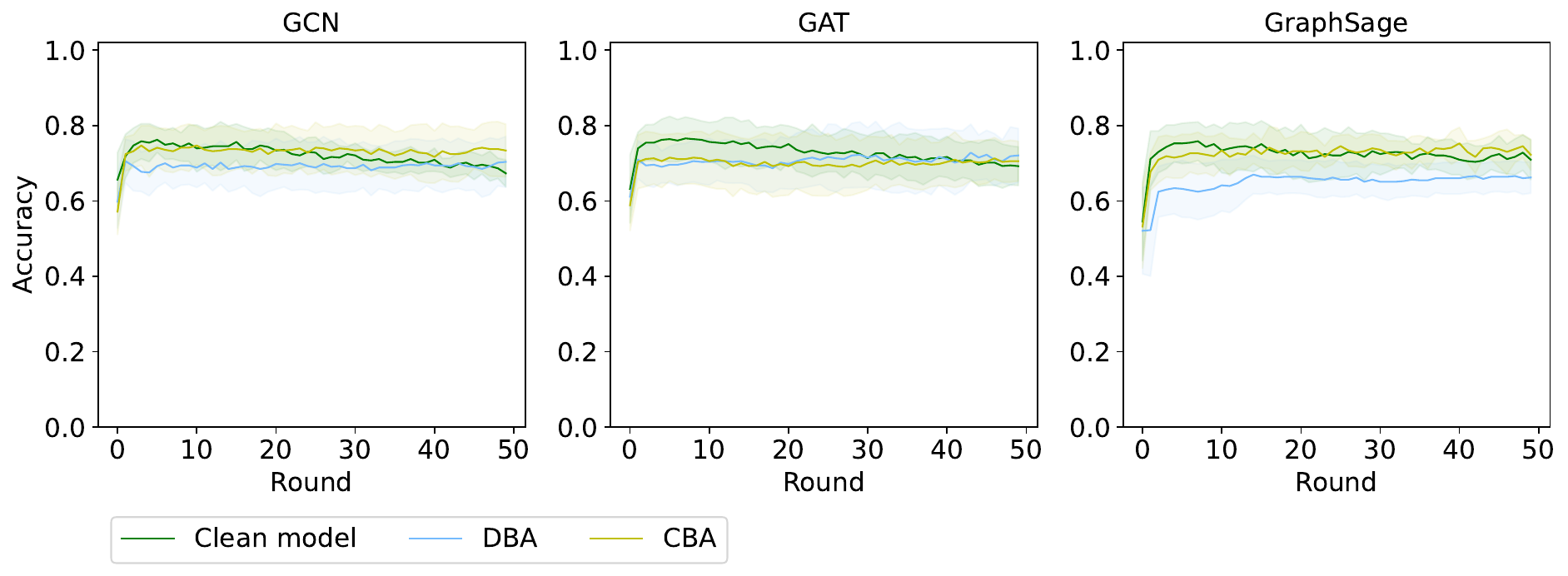}
         \caption{Testing accuracy}
         \label{fig:proteins_defense_asr_honest_majority_b}
     \end{subfigure}
     \caption{Attack success rate on PROTEINS\_full on FoolsGold (in the honest majority attack scenario).}
     \label{fig:proteins_defense_asr_honest_majority}
\end{figure}

The ASR under FoolsGold on NCI1 and PROTEINS\_full datasets (honest majority attack scenario) are shown in Figures~\ref{fig:NCI1_defense_asr_honest_majority} and~\ref{fig:proteins_defense_asr_honest_majority}, respectively. There is a slight increase in the attack success rate of DBA and CBA under FoolsGold, which indicates that this defense fails to identify the malicious updates and misclassifies them as benign. The graph data are not Euclidean data, e.g., images, so the slightly different subgraphs used as triggers do not induce aligned updates. As a result, the cosine similarity cannot be used to detect malicious clients based on their updates. Even though there are more malicious clients in the malicious majority scenario and the probability of detecting the malicious updates should be higher, we observe the same behavior. This further verifies our hypothesis that the defense based on cosine similarity between updates is not very effective in the graph domain.
The clean accuracy drop under FoolsGold on these two datasets is similar to that without the defense. Thus, the defense does not affect the original task in that case.

\section{Impact of Percentage of Malicious Clients}
\label{sec:appendix-impact-asr-m}
\begin{figure}
    \centering
    \includegraphics[width=0.48\textwidth]{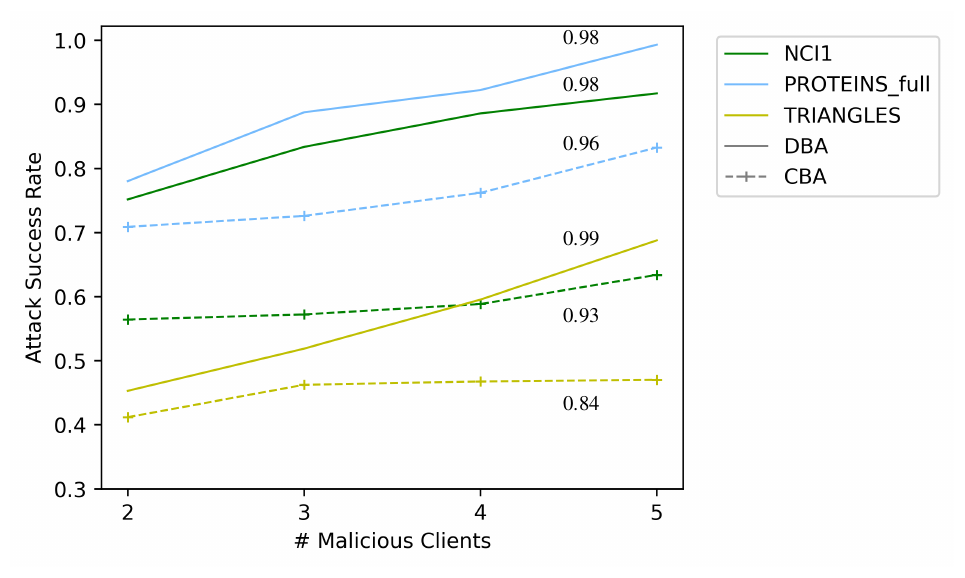}
    \caption{Correlation between ASR and $M$.}
    \label{fig:NM}
\end{figure}

In Figure~\ref{fig:NM}, we show the Pearson Correlation Coefficient of the percentage of malicious clients on the attack performance.

\begin{table*}[!h]
\footnotesize
 \centering
 \caption{FoolsGold weight in DBA and CBA on TRIANGLES (honest majority attack scenario).}
\begin{tabular}{c|c|c|c|c|c|c} 
 \hline
 Attacks & Attacker 1 & Attacker 2 (client 2 in CBA) & Client 3 & Client 4 & Client 5 & Attackers (sum)\\
 \hline
 \hline
 DBA & $1.00\pm0.00$ & $1.00\pm0.00$ & $0.82\pm0.15$ & $0.81\pm0.09$ & $0.78\pm0.12$ & $2.00\pm0.00$ \\
 \hline
 CBA & $1.00\pm0.00$ & $0.00\pm0.00$ & $0.00\pm0.00$ & $0.00\pm0.00$ & $0.00\pm0.00$ & $1.00\pm0.00$ \\
 \hline
\end{tabular}
\label{table:foolsgold_weight}
\end{table*}

\section{FoolsGold Weights}
\label{sec:appendix-C}

To verify our hypothesis (Section~\ref{sec:defense}) for a reason behind the attack performance of DBA and CBA against the FoolsGold defense, we reported the FoolsGold weights on every client in the DBA and CBA on the GraphSage model, as shown in Table~\ref{table:foolsgold_weight}. Here, the FoolsGold weight for each client ranges from $0$ to $1$. 
As we can see, in CBA, the weight of the malicious client is $1$, and the weights of other clients are $0$, which means only the malicious updates are aggregated into the global model. Therefore, the attack success rate of CBA increases significantly under FoolsGold.

On the other hand, in DBA, the weights of the malicious clients are similar to the honest clients, indicating that the malicious and honest updates contribute equally to the aggregated model, as in the aggregation function without the defense, i.e., the average aggregation function. Therefore, there is no obvious difference between the attack performance of DBA before and after the defense. 
The reported weights in Table~\ref{table:foolsgold_weight} verify that our hypothesis is valid.


\end{document}